\documentclass{ws-rv961x669}
\usepackage{ws-rv-van}     
\usepackage{ws-rv-thm}     
\usepackage{subfigure}     
\makeindex

\usepackage{physics}
\usepackage{siunitx}

\usepackage{color,soul}
\usepackage{xspace}
\usepackage{xcolor}

\newcommand{\red}[1]{\textcolor{black}{#1}}

\renewcommand{\vec}[1]{\ensuremath{\boldsymbol{#1}}}

\begin{document}

\chapter[Critical Fluctuations in Polymer Solutions]{Critical Fluctuations in Polymer Solutions: \\ Crossover from Criticality to Tricriticality \label{ra_ch}}

\author[Mikhail A. Anisimov, Thomas J. Longo, and Jan V. Sengers]{Mikhail A. Anisimov, Thomas J. Longo, and Jan V. Sengers}


\address{Institute for Physical Science and Technology \\ and Department of Chemical and Biomolecular Engineering, \\ University of Maryland, College Park, MD 20742, USA}

\begin{abstract}
Critical fluctuations in fluids and fluid mixtures yield a nonanalytic asymptotic Ising-like critical thermodynamic behavior in terms of power laws with universal exponents. In polymer solutions, the amplitudes of these power laws depend on the degree of polymerization. Nonasymptotic behavior (upon the departure from the critical point) is particularly interesting in the case of polymer solutions, where it is governed by a competition between the correlation length of the critical fluctuations and the radius of gyration of the polymer molecules. If the correlation length is the dominant length scale, Ising-like critical behavior is observed. If, however, the radius of gyration exceeds the correlation length, tricritical behavior with mean-field critical exponents is observed. The Ising-like critical region shrinks with the increase of the polymer molecular weight. In the limit of an infinite degree of polymerization, the Ising-like critical region vanishes, yielding to theta-point tricriticality. 

\end{abstract}


\body


\section{Introduction}
\begin{figure}[t]
	\centering
	\includegraphics[width=\textwidth]{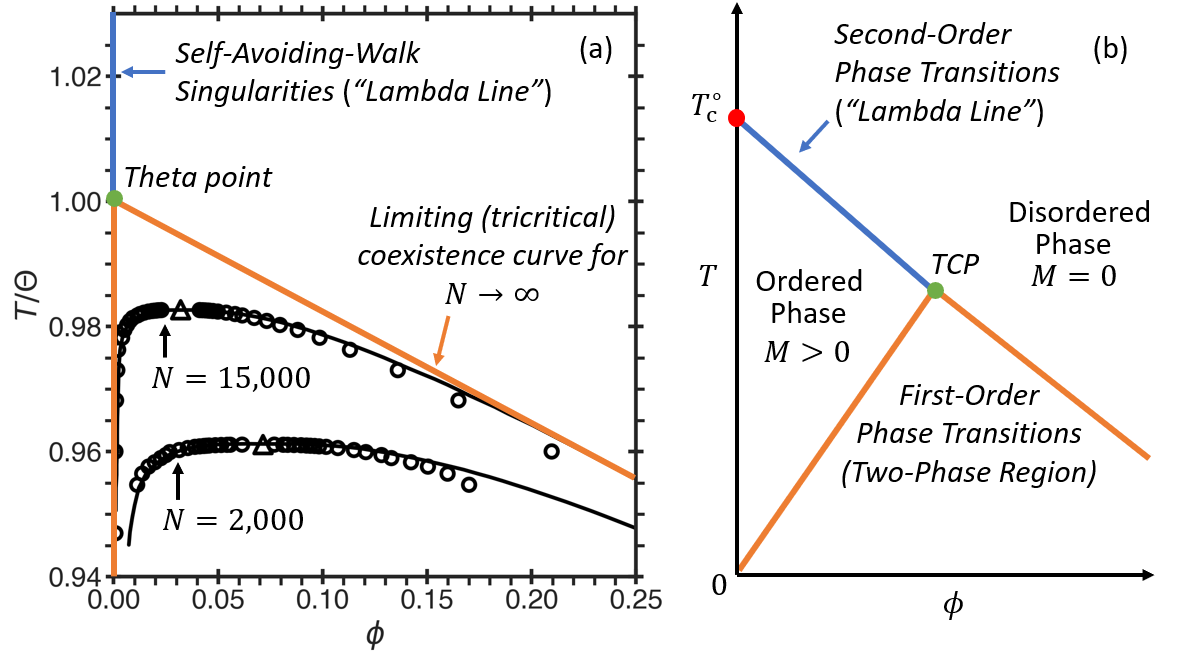}
	\caption{Comparison of the theta-point phase behavior of a polymer solution in the limit of an infinite degree of polymerization ($N$) with a ferromagnet containing a non-magnetic impurity. (a) Phase behavior of polymer solutions of polystyrene in cyclohexane. As the polymer chain becomes longer, the curved shape of the liquid-liquid coexistence transforms into the ``angle-like'' shape near the theta point, where the two branches of the phase coexistence become pure solvent and a phase containing the polymer (volume fraction, $\phi>0$)\cite{Povodyrev_Crossover_1999}. The open circles are experimental data, \cite{Kojima_Coex_1975,Nakata_Coex2_1975,Nakata_Coex1_1978} the triangles are the critical points of demixing, and the solid black curves are the theoretical prediction \cite{Hager_Crossover_2002}. (b) Schematic phase diagram for a ferromagnet (with magnetization, $M$) containing a non-magnetic impurity (with concentration, $\phi$). $T_\text{c}^{\circ}$ (red circle) is the Curie temperature of  a pure ferromagnet. The tricritical point (TCP) separates the line of second-order phase transitions from the first-order phase transitions, accompanied by phase separation.}
	\label{Fig_Tricrit_Comp}
\end{figure}

Almost thirty years ago, Michael E. Fisher published a review on phase transitions in ionic fluids \cite{Fisher_Coulombic_1994}, in which he discussed a remarkable analogy between the demixing of polymer solutions at an infinite degree of polymerization and a tricritical phase transition, as observed in a variety of systems, such as He$^3$-He$^4$ mixtures \cite{Vollhardt_He_1990,Schmitt_He_2015} and metamagnets \cite{Jacobs_Metamagnet_1967,Fisher_Magnetic_1975}, with coupled order parameters \cite{Anisimov_Coupling_1981,Knobler_PT_1984,anisimov_critical_1991}.  We illustrate this analogy in Fig.~\ref{Fig_Tricrit_Comp} by comparing the phase behavior in solutions of polystyrene in cyclohexane with a ferromagnet containing a non-magnetic impurity. In this paper, we describe the physics that governs the crossover between Ising-like asymptotic critical behavior and tricritical theta-point behavior in high-molecular-weight polymer solutions.

Thermodynamic behavior of substances asymptotically close to their critical points (second-order phase transitions), follows the principle of critical-point universality: the microscopic details become unimportant if the correlation length of the order-parameter fluctuations exceeds the range of intermolecular interactions \cite{Stanley_PT_1971,Wilson_RG_1976,fisher_scaling_1983,wilson_renormalization_1983,Goldenfeld_Lectures_1992,Anisimov_CriticalRegion_2000,Pelissetto_RGReview_2002}. All fluids and fluid mixtures, simple and complex, belong to the three-dimensional Ising-model universality class \cite{fisher_scaling_1983,Anisimov_CriticalRegion_2000}. Asymptotically, their thermodynamic properties are described by power laws with universal Ising-model critical exponents and system-dependent amplitudes (which are interrelated by universal relations)\cite{Pelissetto_RGReview_2002}. However, there is a famous quotation of Peter Debye in response to the critical power laws (proposed by Michael Fisher) at the first international conference on critical phenomena \cite{DebyeQuote}:
\begin{quote}
	``I would like that the theoretical people tell me when I am so and so far away from the critical point, then my curve should look so and so.''
\end{quote} 
This question received the answer in the theory of crossover critical phenomena developed in the 1980s and 1990s \cite{Nicoll_Crossover_1985,Chen_Crossover1_1990,Chen_Crossover2_1990,Tang_Crossover_1991,Kostrowicka_Crossover_1999,Povodyrev_Crossover_1999,Kim_Crossover_2003}. The crossover between asymptotic Ising critical behavior and mean-field behavior away from the critical point is governed by a ratio of the distance between the molecules and the range of molecular interactions, known as the Ginzburg number \cite{Anisimov_Crossover_Gen_1992,Anisimov_CriticalRegion_2000}. Since, in ordinary fluids, the range of molecular interactions does not exceed the intermolecular distance, there is no region where the critical anomalies can be fully described by the mean-field (``Landau'') theory \cite{LL_Stat_Phys}. The only possibility to observe complete crossover from Ising criticality to mean-field criticality in simple systems is to consider an Ising-like model with a tunable range of interactions. An accurate computational study of such a model \cite{Kim_Crossover_2003} \red{has} perfectly confirmed the crossover theory.

An alternative kind of crossover from Ising criticality to mean-field-like behavior, the subject of this review, is a striking phenomenon observed in solutions of high-molecular-weight polymers in a low-molecular-weight solvent. At a finite degree of polymerization, asymptotically close to the demixing critical point, one observes universal Ising criticality where the amplitudes of the power laws depend on the degree of polymerization \cite{deGennes_CritOpal_1968,deGennes_Book_1979,Sanchez_Scaling_1989}. Upon an increase in the degree of polymerization, the Ising-like critical region shrinks, yielding to theta-point tricritical behavior \cite{Povodyrev_Crossover_1999,Agayan_Cross_2000,Hager_Crossover_2002,Kostko_Crossover_2002,Kostko_Crossover_2005,Anisimov_PolyTricriticality_2005}. The crossover to theta-point tricriticality is governed by a competition between the correlation length of the critical fluctuations and the radius of gyration of the polymer molecules. If the correlation length is the dominant length scale, Ising-like critical behavior is observed. If, however, the radius of gyration exceeds the correlation length, tricritical behavior is observed \cite{Kostko_Crossover_2005,Anisimov_PolyTricriticality_2005}. \red{Theta}-point tricritical behavior is characterized by mean-field critical exponents \cite{Knobler_PT_1984,anisimov_critical_1991}, but the tricritical amplitudes are affected by fluctuation-induced logarithmic corrections \cite{Hager_Crossover_2002}. This picture of global non-asymptotic critical behavior in polymer solutions has been confirmed by light-scattering experiments in our laboratory \cite{Jacob_Scattering_2001,Kostko_Crossover_2002,Kostko_Crossover_2005,Kostko_Diffusion_2007}. 

This chapter is organized as follows. In Section~\ref{Sec2_CritPhenom}, we discuss the Ising-like critical phenomena in polymer solutions at finite degrees of polymerization. In particular, we show that accurate experiments validate the de Gennes-Sanchez scaling for the critical amplitudes \cite{deGennes_CritOpal_1968,deGennes_Book_1979,Sanchez_Scaling_1989}. In Section~\ref{Sec1_MFTricrit}, we review the thermodynamics of the theta-point in polymer solutions and show that the theta point is a tricritical point. In Section~\ref{Sec3_Crossover}, we demonstrate the crossover between Ising criticality and theta-point tricriticality in the shape of the liquid-liquid coexistence and the values of the critical exponents. In Section~\ref{Sec_Polymer_Blend}, we discuss the Ising-like critical phenomena in polymer blends. Lastly, in Section~\ref{Sec_Conclusion}, we \red{present} some concluding remarks.

\section{Ising-Like Critical Phenomena in Polymer Solutions \label{Sec2_CritPhenom}}
\begin{figure}[t]
	\centering
	\includegraphics[width=0.49\textwidth]{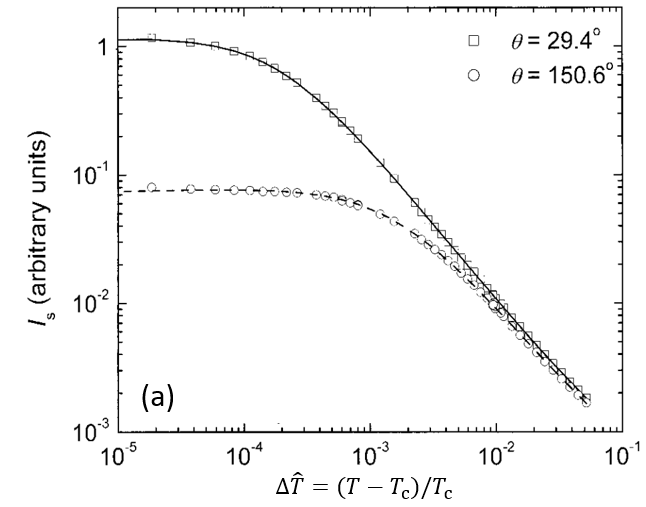}
	\includegraphics[width=0.49\textwidth]{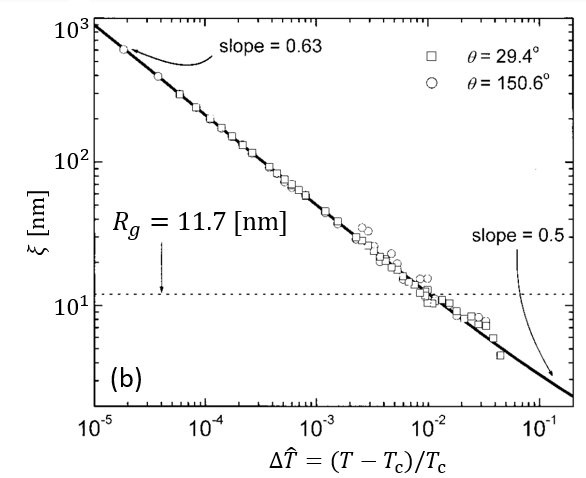}
	\caption{Demonstration of Ising-model scaling behavior in polymer solutions from. (a) Light-scattering intensity for a solution of polystyrene ($M_\text{w} = 195,900 \text{ g/mol}$) in cyclohexane as a function of the distance to the critical temperature. (b) Correlation length for the same system. The symbols represent the experimental data \cite{Jacob_Scattering_2001}, while the solid curves represents the theoretical prediction \cite{Jacob_Scattering_2001,Kostko_Crossover_2005}. Adapted with permission from Ref.~\cite{Jacob_Scattering_2001} \copyright~The Optical Society.}
	\label{Fig_Jabi}
\end{figure}

The thermodynamic behavior of polymer solutions near their critical demixing transitions, like of all fluids and fluid mixtures, belongs to the three-dimensional Ising-model universality class. Asymptotically close to the liquid-liquid critical point, the thermodynamic properties of polymer solutions (\textit{e.g.}, phase coexistence, osmotic \red{susceptibility}, correlation length, \textit{etc.}) are described by power laws with universal critical exponents, as given in Table~\ref{Table_CritExp}. This scaling behavior has been confirmed in a variety of systems, including polystyrene in cyclohexane \cite{Kojima_Coex_1975,Nakata_Coex2_1975,Nakata_Coex1_1978,Jacob_Scattering_2001,Hager_Crossover_2002,Kostko_Crossover_2002,Kostko_Crossover_2005,Anisimov_PolyTricriticality_2005,Kostko_Diffusion_2007}, polystyrene in methylcyclohexane \cite{Dobashi_MethPS_1980}, and in polymethylmethacrylate in 3-octanone \cite{Xia_PMMA_1992,Xia_PMMA_1996}. In this Section, we focus on light-scattering measurements in solutions of polystyrene in cyclohexane. For instance, as shown in Fig.~\ref{Fig_Jabi}, accurate light-scattering measurements in a solution of polystyrene ($M_\text{w} = 195,900 \text{ g/mol}$) in cyclohexane, performed by Jacob \textit{et al.}\cite{Jacob_Scattering_2001}, have confirmed the universal Ising-model behavior of the correlation length, $\xi$, and of the osmotic susceptibility, $\hat{\chi}$.

In light-scattering experiments on near-critical mixtures, the intensity is proportional to the spatially-dependent (``local'') osmotic susceptibility, $\hat{\chi}(q\xi)$, and can be written as\cite{Kostko_Crossover_2002,Kostko_Crossover_2005}
\begin{equation}
	I_\text{s} = I_0\hat{\chi}(q\xi) + I_\text{b}\text{,}
\end{equation}
where $q = 4\pi n \lambda_0^{-1} \sin(\theta/2)$ is the light-scattering wave number, in which $n$ is the refractive index\red{,} $\lambda_0$ is the wavelength of incident light\red{, and $\theta$ is the scattering angle}, while $I_0$ is an instrumental constant and $I_\text{b}$ is a regular ``background'' scattering intensity (not associated with critical fluctuations). In the mean-field approximation and in the first-epsilon expansion of the Renormalization Group theory (RG) \cite{Wilson_Fisher_1972,Agayan_Crossover_2001}, the susceptibility is described through the Ornstein-Zernike correlation function \cite{LL_Stat_Phys}
\begin{equation}\label{Eq_OrnZernike}
	\hat{\chi} = \frac{\hat{\chi}_{q=0}}{1+q^2\xi^2}\text{,}
\end{equation}
where $\hat{\chi}_{q=0}$ is the ``bulk'' (thermodynamic) susceptibility. In the mean-field approximation and in the first-epsilon RG expansion, the correlation length is related to the thermodynamic susceptibility as $\xi = (c_0\hat{\chi}_{q=0})^{1/2}$, where $c_0$ is \red{of} the order of the square of the range of molecular interactions \cite{LL_Stat_Phys}. 

\begin{table}[t]
	\tbl{\red{Asymptotic} power laws of thermodynamic properties  with the reduced distance to the critical temperature, $\Delta\hat{T}=(T-T_\text{c})/T_\text{c}$, and their critical exponents.\cite{Sengers_Mesothermo_2010} The asymptotic critical amplitudes of each property ($A_0$, $B_0$, $\Gamma_0$, $\xi_0$, and $\sigma_0$) are system dependent. \red{In} the text, we use an upper bar to differentiate between the mean-field and scaling theory prediction for the amplitude.}{
		\begin{tabular}{@{}lccc@{}}
			\toprule\\[-12pt]
			& 			   & Mean-Field			& $3d$ Ising-Model   \\
			Property			&  Scaling Law & Critical Exponents & Critical Exponents \\ [3pt]
			\hline\\[-6pt]
			Heat Capacity		& $C = A_0 |\Delta\hat{T}|^{-\alpha}$ 	         & $\alpha = 0.0$  & $\alpha = 0.110\pm 0.003$ \\[3.5pt]
			Order Parameter & $\varphi = \pm B_0 |\Delta\hat{T}|^{\beta}$       & $\beta = 0.5$   & $\beta = 0.326\pm 0.002$ \\[3.5pt]
			Susceptibility		& $\hat{\chi} = \Gamma_0 |\Delta\hat{T}|^{-\gamma}$    & $\gamma = 1.0$  & $\gamma = 1.239\pm 0.002$\\[3.5pt]
			Correlation Length  & $\xi = \xi_0 |\Delta\hat{T}|^{-\nu}$				 & $\nu = 0.5$	   & $\nu = 0.630\pm 0.002$   \\[3.5pt]
			Interfacial Tension & $\sigma = \sigma_0 |\Delta\hat{T}|^{\mu}$			 & $\mu = 1.5$	   & $\mu = 1.260 \pm 0.004$   \\[3pt]
			\Hline
	\end{tabular}}
	\label{Table_CritExp}
\end{table}

In the scaling theory, the \red{spatially dependent} susceptibility is best described by using the following interpolation based on the Fisher-Burford approximation \cite{Fisher_Corr_1964,Burford_Approx_1967},
\begin{equation}\label{Eq_ScalingSuscept}
	\hat{\chi} \simeq  \frac{\hat{\chi}_{q=0}}{(1+q^2\xi^2)^{1-\eta/2}}\text{.}
\end{equation}
In the scaling theory, the thermodynamic susceptibility is related to the correlation length by $\hat{\chi}_{q=0}\sim \xi^{2-\eta}$. Thus, the critical exponent of the susceptibility, $\gamma$, is universally connected to that of the correlation length, $\nu$, through the relation $\gamma = \nu(2-\eta)$ \cite{fisher_scaling_1983,wilson_renormalization_1983,Goldenfeld_Lectures_1992,Anisimov_CriticalRegion_2000,Pelissetto_RGReview_2002}. The difference between Eqs.~(\ref{Eq_OrnZernike}) and (\ref{Eq_ScalingSuscept}) is almost negligible because the critical exponent, $\eta$, in the Ising-model universality class is a very small number, $\eta = 0.033$ \cite{Halperin_Divergent_1974,Siggia_dynamics_1976,Sengers_Transport_1985}. The two limits of the susceptibility, $\hat{\chi}(q\xi\ll 1)\sim \xi^{2-\eta}$ and $\hat{\chi}(q\xi\gg 1)\sim q^{\eta-2}$, have been observed in the light-scattering measurements of Jacob \textit{et al.}\cite{Jacob_Scattering_2001}, presented in Fig.~\ref{Fig_Jabi}a, and in Anisimov \textit{et al}\cite{Kostko_Crossover_2005}. In addition, it was observed that the correlation length, as shown in Fig.~\ref{Fig_Jabi}b, crosses over from asymptotic scaling behavior toward mean-field behavior far away from the critical point. This occurs when the correlation length becomes smaller than the radius of gyration. Because of two scaling relations between critical exponents, namely, $d\nu = 2-\alpha$, where $d$ is the dimensionality of space, and $2-\alpha = 2\beta +\gamma$ \cite{fisher_scaling_1983,wilson_renormalization_1983,Goldenfeld_Lectures_1992,Anisimov_CriticalRegion_2000,Pelissetto_RGReview_2002}, one can see that by measuring the critical exponents of any two thermodynamic properties all other exponents become known.

\begin{figure}[t]
	\centering
	\includegraphics[width=0.48\textwidth]{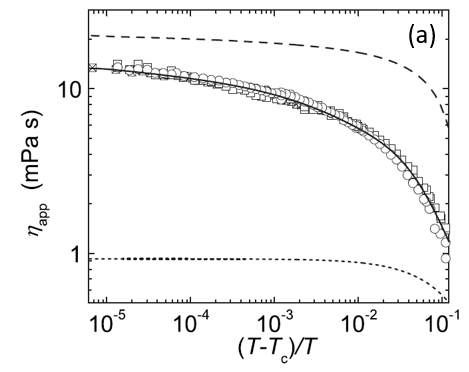}
	\includegraphics[width=0.50\textwidth]{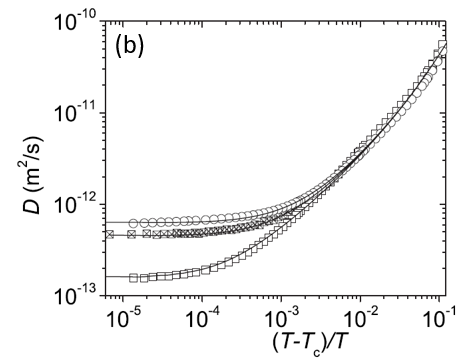}
	\caption{(a) Apparent (mesoscopic) viscosity of a solution of polystyrene ($M_\text{w} = 195,900 \text{ g/mol}$) in cyclohexane. The data correspond to three light-scattering angles: $\theta =\SI{30}{\degree}$ (open squares), $\theta=\SI{90}{\degree}$ (crossed squares), and $\theta = \SI{150}{\degree}$ (open circles). The solid curve is the prediction from the mode-coupling theory, the dashed curve represents the viscosity of the solution, and the dotted curve represents the viscosity of the solvent. (b) Diffusion coefficient for the same system. The curves represent the fit of the mode-coupling theory to the experimental data using the mesoscopic viscosity presented in (a) \cite{Kostko_Diffusion_2007}. Reprinted with permission from \textit{Phys. Rev. E} \cite{Kostko_Diffusion_2007}.}
	\label{Fig_PolyDiff2007}
\end{figure}

A further confirmation of the scaling behavior of polymer solutions has been observed through the measurements of the mutual diffusion coefficient, $D$, by dynamic light scattering \cite{Jacob_Scattering_2001,Kostko_Diffusion_2007}. The diffusion coefficient may be expressed as a sum of a critical contribution, $D_\text{c}$, and a background contribution, $D_\text{b}$, as $D=D_\text{c}+D_\text{b}$. The mode-coupling theory of critical dynamics predicts that, asymptotically close to the critical point, $D_\text{c}$, should satisfy a modified Stokes-Einstein-Kawasaki relation of the form \cite{Halperin_Divergent_1974,Kawasaki_1976,Burstyn_Dynamic_1983,Kostko_Diffusion_2007}
\begin{equation}\label{Eq_CritDiff}
	D_\text{c} = \frac{R_\text{D}k_\text{B}T}{6\pi\eta_\text{app}\xi}K(q\xi)\left[1 + \left(\frac{q\xi}{2}\right)^2\right]^{z_\eta/2}\Omega(\xi/\xi_\text{D})\text{,}
\end{equation}
where $\eta_\text{app}$ is the apparent mesoscopic viscosity, $k_\text{B}$ is Boltzmann's constant, and $R_\text{D}$ is a universal dynamic amplitude ratio, typically, of the order of unity \cite{Kawasaki_1976,Kostko_Diffusion_2007}. Asymptotically close to the critical temperature, the actual shear viscosity of the solution\red{, $\eta$ (not to be confused with the critical exponent of the correlation function),} diverges as $\eta = \eta_\text{b}(\red{Q_0}\xi)^{z_\eta}$, in which $\red{Q_0}$ is a system-dependent amplitude, $z_\eta = 0.068$ \cite{Hao_Viscosity_2005}, and $\eta_\text{b}$ is the noncritical background viscosity \cite{Ohta_Viscosity_1977}. Figure~\ref{Fig_PolyDiff2007}a depicts the behavior of the viscosity in a solution of polystyrene in cyclohexane. In Eq.~(\ref{Eq_CritDiff}), the factor in square brackets provides the necessary crossover from the diverging critical viscosity to the constant background viscosity far away from the critical point, $q\xi\gg 1$. Meanwhile, the Kawasaki function, $K(q\xi)\equiv K(x) = [3/(4x^2)](1+x^2+[x^3-x^{-1}]\arctan{x})$, satisfies the limits: $K(q\xi\ll 1) = 1$ and $K(q\xi\gg 1)=q\xi$ \cite{Kawasaki_1976}. Lastly, the factor $\Omega(\xi/\xi_\text{D}) = (2/\pi)\arctan(\xi/\xi_\text{D})$ accounts for deviations from asymptotic dynamic critical behavior due to a finite cut-off length scale, $\xi_\text{D}$, that is of the order of the radius of gyration. It follows from Eq.~(\ref{Eq_CritDiff}), that the diffusion coefficient vanishes asymptotically as $D_\text{c}\sim\xi^{-(1+z_\eta)}$ at the critical point. The background contribution to the diffusion coefficient, $D_\text{b}$, can be estimated as\cite{Kostko_Diffusion_2007}
\begin{equation}\label{Eq_DiffBack}
	D_\text{b} \approx \frac{k_\text{B}T}{16\eta_\text{b}\xi}\left[\frac{1+q^2\xi^2}{\red{q_\text{C}}\xi}\right]\text{\red{,}}
\end{equation}
\red{where the wavenumber, $q_\text{C}$, can be approximated as $\xi_\text{D}^{-1}$ \cite{Kostko_Diffusion_2007}.} From Eqs.~(\ref{Eq_CritDiff}) and (\ref{Eq_DiffBack}), the behavior of the diffusion coefficient has the following two limits. \red{In the hydrodynamic regime ($q\xi\ll 1$), the diffusion coefficient becomes independent of $q$ varying with the distance to the critical temperature as $D\sim |\Delta\hat{T}|^{\nu-z_\eta}$ \cite{Onuki_2002,Sengers_2006}, while in the non-hydrodynamic limit ($q\xi\gg 1$), $D(q)$ becomes independent of the temperature.} Dynamic light-scattering measurements of Kostko \textit{et al.}\cite{Kostko_Diffusion_2007}, presented in Fig.~\ref{Fig_PolyDiff2007}b, have confirmed this behavior of the diffusion coefficient.

\begin{figure}[t]
	\centering
	\includegraphics[width=\textwidth]{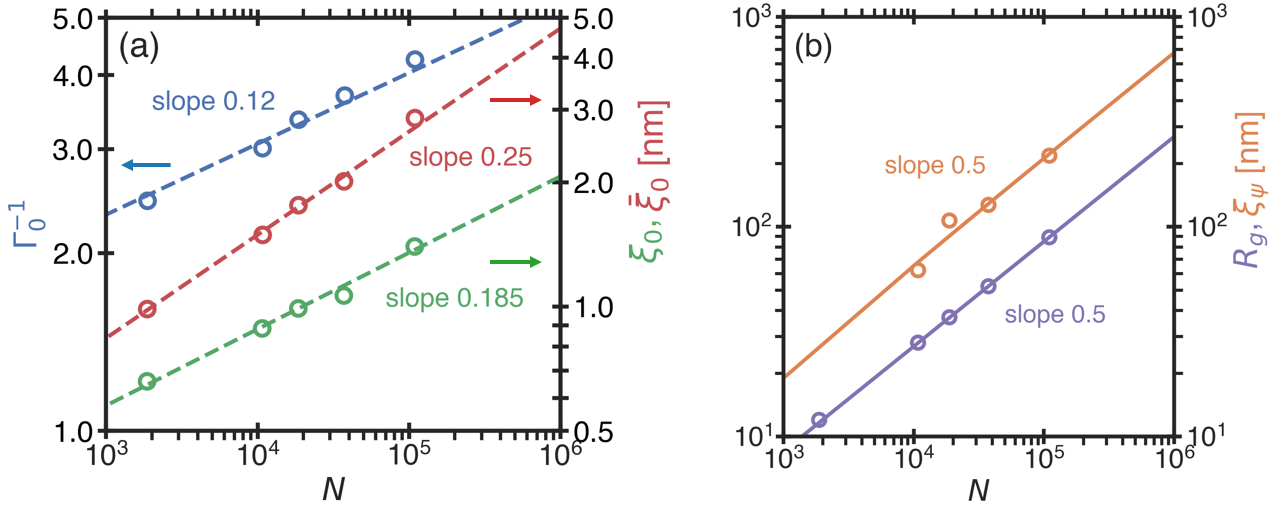}
	\caption{(a) Experimental confirmation of the de Gennes-Sanchez scaling relations (Table~\ref{Table_deGenSanchez}). The left axis presents the amplitude of the susceptibility, $\Gamma_0$ (blue), while the right axis presents the amplitudes of the correlation length of concentration fluctuations, in the critical region, $\xi_0$ (blue), and away from the critical point, $\bar{\xi}_0$, in the mean-field region (orange) \cite{Kostko_Crossover_2005}. The slopes of the dashed lines correspond to the values predicted by de Gennes and Sanchez, as given in Table~\ref{Table_deGenSanchez}. (b) The radius of gyration $R_\text{g}$, as a function of the degree of polymerization $N$, from neutron-scattering measurements \cite{Melnichenko_Neutron1_1997,Melnichenko_Neutron2_2005}. The purple line corresponds to the scaling law, $R_\text{g}\sim N^{1/2}$, along the critical line. Also depicted is the correlation length associated with the polymer-chain fluctuations, $\xi_\psi$, obtained from the coupling between diffusive and viscoelastic dynamic modes \cite{Kostko_Diffusion_2007}. The orange line corresponds to the theoretical prediction $\xi_\psi\propto R_\text{g}$ for large $N$ - more details provided in Sec.~\ref{Sec3_Crossover}, see Eq.~(\ref{Eq_Ch5_Corr_PolyChain}).}.
	\label{Fig_Light_Scat_Amp}
\end{figure}

While the critical exponents in polymer solutions, like in all fluids and fluid mixtures, belong to the Ising-model universality class, the critical amplitudes, unique to polymer solutions, diverge or vanish with the degree of polymerization, $N$, as given in Table~\ref{Table_deGenSanchez}. Static light-scattering measurements of solutions of polystyrene in cyclohexane from $M_\text{w} = \SI{195,900}{\gram/\mole}$ to $M_\text{w} = 11.4$ million $\si{\gram/\mole}$, \red{performed in our laboratory} \cite{Kostko_Crossover_2005}, have confirmed that the amplitude of the correlation length and \red{the amplitude of the susceptibility} satisfy the predictions \red{of de Gennes and Sanchez} \cite{deGennes_CritOpal_1968,deGennes_Book_1979,Sanchez_Scaling_1989}, \red{both in the vicinity of the critical point and in the mean-field regime far away from the critical point and in the mean-field regime far away from the critical point} (see Fig.~\ref{Fig_Light_Scat_Amp}). The $N$-dependent nature of the amplitudes raises a fundamental question about the critical behavior of extremely high-molecular-weight polymer solutions close to the limit of an infinite degree of polymerization.

\begin{table}[t]
	\tbl{de Gennes-Sanchez scaling relations for various thermodynamic properties in polymer solutions \cite{deGennes_CritOpal_1968,deGennes_Book_1979,Sanchez_Scaling_1989}. The amplitude exponents are given by their mean-field and three-dimensional Ising-model forms.}
	{\begin{tabular}{@{}lcccc@{}}
			\toprule\\[-12pt]
			&															   & de Gennes-Sanchez & Mean-Field					& Ising-Model  \\ 
			Property              & Scaling Law                                                & Relations          & Amplitude & Amplitude \\[3pt]
			\hline\\[-6pt]
			Order Parameter   & $\varphi\sim \pm {N}^{-b}|\Delta\hat{T}|^{\beta}$   & $b=(1-\beta)/2$    & $b=0.25$ & $b=0.34$ \\[3.5pt]
			Heat Capacity         & $C \sim {N}^{-a}|\Delta\hat{T}|^{-\alpha}$         & $a = (1+\alpha)/2$ & $a=0.5$  & $a=0.555$ \\[3.5pt]
			Susceptibility        & $\hat{\chi} \sim {N}^{g}|\Delta\hat{T}|^{-\gamma}$ & $g = (1-\gamma)/2$ & $g=0$    & $g=-0.12$ \\[3.5pt]
			Correlation Length    & $\xi \sim {N}^{n}|\Delta\hat{T}|^{-\nu}$           & $n = (1-\nu)/2$    & $n=0.25$ & $n=0.185$ \\[3.5pt]
			Interfacial tension   & $\sigma\sim {N}^{-m}|\Delta\hat{T}|^{\mu}$         & $m = (2-\mu)/2$    & $m=0.25$ & $m=0.37$ \\[3pt]    
			\Hline
	\end{tabular}}\label{Table_deGenSanchez}
\end{table}

Typically, the critical behavior of simple fluids is controlled by a single mesoscopic length scale, the correlation length of the concentration fluctuations, $\xi$ \cite{fisher_scaling_1983,Goldenfeld_Lectures_1992,Anisimov_CriticalRegion_2000,Pelissetto_RGReview_2002}. However, in polymer solutions there is a second competing length scale, namely, the size of the polymer coil (as defined through the radius of gyration, $R_\text{g}$) that depends on the degree of polymerization, $N$. With increasing values of $N$, upon approaching the limit $N\to\infty$, the critical concentration (volume fraction), $\phi_\text{c}$, decreases to zero, while the critical temperature, $T_\text{c}$, approaches the theta temperature (see Fig.~\ref{Fig_Tricrit_Comp}). In the asymptotic vicinity of the critical temperature, where the correlation length is larger than the radius of gyration, the range of Ising-like critical behavior decreases as $R_\text{g}$ increases. The radius of gyration in the polystyrene-cyclohexane system, obtained by neutron-scattering experiments \cite{Melnichenko_Neutron1_1997,Melnichenko_Neutron2_2005}, obeys the scaling behavior $R_\text{g}\sim N^{1/2}$ along the critical line, \red{see Fig.~\ref{Fig_Light_Scat_Amp}b}. Dynamic light-scattering experiments reveal a second correlation length \cite{Kostko_Diffusion_2007}, associated with the entanglements and disentanglements of the polymer chain (a ``viscoelastic'' length scale\cite{Tanaka_Dynamic_2002}), which we identify with $\xi_\psi$. \red{Figure~\ref{Fig_Light_Scat_Amp}b verifies} the theoretical prediction $\xi_\psi\propto R_\text{g}$ (see the next Section for details). Thus, the asymptotic amplitude of the correlation length scales with the degree of polymerization as $\xi_0\sim N^{0.185}$ in the Ising-like critical region (where $\xi > \xi_\psi$), while it scales as $\bar{\xi}_0\sim N^{0.25}$ in the mean-field region (where $\xi < \xi_\psi$), see Fig.~\ref{Fig_PolyDiff2007}a. Such shrinking of the scaling critical regime is also observed in the behavior of the coexistence curve (see Fig.~\ref{Fig_Tricrit_Comp}), which leads to the ``angle-like'' shape in the limit of an infinite degree of polymerization. In the next Section, we prove thermodynamically that in the limit $N\to\infty$ the Ising-critical regime disappears completely and the polymer solution exhibits theta-point tricritical behavior.

\section{Tricritical Behavior near the Theta-Point \label{Sec1_MFTricrit}}
The phase separation behavior of a polymer solution below the critical point of demixing was originally explained in the mean-field theory by Flory and Huggins \cite{Flory_Polymer_1941,Huggins_Solutions_1941,Flory_Theory_1953}. The Flory-Huggins theory, based on a lattice model, in which the monomers link to form linear polymer chains and are assumed to be the same size as the solvent molecules. According to the Flory-Huggins theory\cite{Flory_Theory_1953}, the Gibbs free energy of mixing per monomer unit is given by
\begin{equation}\label{Eq_FloryFreeEng}
	\frac{\Delta G}{n \red{k_\text{B}}T}\equiv \hat{G} = (1-\phi)\ln(1-\phi) + \frac{\phi}{N}\ln\phi + \varpi\phi(1-\phi)\text{,}
\end{equation}
where $n$ is the total number of solvent and monomer molecules, $\phi$ is the volume fraction (the monomer-solvent number ratio), and $\varpi=\Theta/2T$ is the Flory interaction parameter (where $\Theta$ is known as the ``theta'' temperature). We consider the case when the mixture is incompressible and $\Theta$ is independent of pressure. The theta temperature is the limiting value of the critical temperature of phase separation in the limit of an infinite degree of polymerization \cite{Colby_Theta_1990,Teixeira_Theta_1999,Rubinstein_Polymer_2003}.

The critical demixing point is found from the thermodynamic stability conditions, $(\partial^2\hat{G}/\partial\phi^2)|_{T,N}=0$ and $(\partial^3\hat{G}/\partial\phi^3)|_{T,N}=0$. In the Flory-Huggins theory, the critical temperature, $T_\text{c}$, and critical volume fraction, $\phi_\text{c}$, of demixing depend on the degree of polymerization, $N$, as \cite{Flory_Theory_1953}
\begin{equation}\label{Eq_FloryCritParams}
	T_\text{c} = \frac{\Theta}{(1+1/\sqrt{N})^2}\quad\quad \text{and} \quad\quad \phi_\text{c} = \frac{1}{1+\sqrt{N}}\text{.}
\end{equation}
 In the limit of $N\to\infty$ in the Flory-Huggins theory, the strong asymmetry in the contribution to the Gibbs energy from the entropy of mixing, causes an ``entropy-driven'' phase transition\cite{Frenkel_Entropy_1999} to occur at the theta point, where $T_\text{c}\to\Theta$ and $\phi_c\to 0$. 
 
Using the relations in Eq.~(\ref{Eq_FloryCritParams}), we define the order parameter as $\varphi = \phi-\phi_\text{c}$, and the reduced distance to the critical temperature as $\Delta\hat{T}=(T-T_\text{c})/T_\text{c}$. Expanding the critical part of the Gibbs free energy in the vicinity of the liquid-liquid critical point and in the large $N$ limit, gives a typical truncated Landau expansion of the form \cite{LL_Stat_Phys,LL_Stat_Phys_II}
\begin{equation}\label{Eq_LndExp_FreeEng}
	\Delta \hat{G} = \hat{G}-\hat{G}_\text{b} = \frac{1}{2!}\Delta\hat{T}\varphi^2 + \frac{1}{4!}u_0\varphi^4\text{,}
\end{equation}
where the coefficient of the fourth-order term is $u_0 = 2\sqrt{N}$ in the large $N$ limit \cite{Povodyrev_Crossover_1999}. The Landau expansion represents mean-field critical behavior. The background (``regular'') part of the Gibbs free energy is given by
\begin{equation}
	\hat{G}_\text{b} = \hat{\mu}_\text{s}(\phi_\text{c}) + \phi\hat{\mu}_\text{ms}(\phi_\text{c},\varpi)\text{,}
\end{equation}
where the chemical potential of the solvent molecules, $\hat{\mu}_\text{s}$, and the monomer-solvent exchange chemical potential, $\hat{\mu}_\text{ms}$, are defined from appropriate derivatives of the Gibbs free energy: $\hat{\mu}_\text{s} = \hat{G} -\phi\hat{\mu}_\text{ms}$ and $\hat{\mu}_\text{ms} = (\partial\hat{G}/\partial\phi)|_{T,N}$. This part of the free energy does not affect either the shape of the phase boundary or the osmotic \red{susceptibility}.

The effects of fluctuations of the order parameter can be incorporated in the mean-field approximation as well as in the first-order RG expansion through the ``local'' (spatial-dependent) Gibbs free-energy density (per unit volume, $V$), known as the Landau-Ginzburg functional\cite{LL_Stat_Phys},
\begin{equation}\label{Eq_FreeFunc}
	\frac{\text{d}(\Delta\hat{G})}{\text{d}V} = \frac{1}{2}\Delta\hat{T}[\varphi(\vec{r})]^2 + \frac{1}{4!}u_0[\varphi(\vec{r})]^4  + \frac{1}{2}c_0|\nabla\varphi(\vec{r})|^2\text{,}
\end{equation}
where $\vec{r}$ is \red{the} spatial coordinate, the first two terms are the ``bulk'' thermodynamic potential, given by the Landau expansion, Eq.~(\ref{Eq_LndExp_FreeEng}), and the last term is the energy penalty associated with the fluctuation-induced inhomogeneous mixing of the solute and solvent. The coefficient, $c_0$, is a length scale on the order of the square of the range of molecular interactions. For polymer solutions, $c_0$, is $N$-dependent, scaling as $c_0\sim\sqrt{N}$ \cite{deGennes_CritOpal_1968,Shinozaki_Amplitudes_1981,Nose_Interfacial_1984,Kostko_Crossover_2005,Mirzaev_HeatCap_2010}. From the Landau-Ginzburg functional, the correlation length is related to the thermodynamic susceptibility through $\xi = \sqrt{c_0\hat{\chi}_{q=0}}$ \cite{LL_Stat_Phys}. Since, in the mean-field approximation, the susceptibility is independent of $N$, scaling as $\hat{\chi}_{q=0} = (\partial^2\Delta\hat{G}/\partial\varphi^2)|_{T,N} \sim 1/\Delta\hat{T}$ from Eq.~(\ref{Eq_LndExp_FreeEng}), then the $N$-dependence of the correlation length emerges from $c_0$, such that $\xi \sim \sqrt{c_0/\Delta\hat{T}} = N^{1/4}|\Delta\hat{T}|^{-1/2}$ (see Table~\ref{Table_deGenSanchez}).

The Landau expansion of the Flory-Huggins free-energy density reveals two problems. First, asymptotically close to the critical point, the Flory-Huggins theory does not capture the Ising-like critical behavior of real solutions (see Table~\ref{Table_CritExp}). For example, the mean-field Flory-Huggins theory predicts that the volume fraction along coexistence, found from $(\partial\Delta\hat{G}/\partial\varphi)|_{T,N}=0$, scales as $\varphi = \pm \sqrt{3}N^{-1/4}|\Delta\hat{T}|^{1/2}$, while asymptotically close to the critical point, $\varphi \sim |\Delta\hat{T}|^\beta$, \red{with} $\beta = 0.326$. Second, Eq.~(\ref{Eq_LndExp_FreeEng}), does not describe the phase behavior in the vicinity of the theta point, where $N\to\infty$. In this limit, the coefficient of the fourth-order term diverges at the theta point, and the Landau expansion no longer remains valid. Therefore, the range of applicability of Eq.~(\ref{Eq_LndExp_FreeEng}) shrinks with increasing degree of polymerization. To \red{solve} these problems, in this Section, we will present a phenomenological thermodynamic theory to describe the theta point as a tricritical point, and, in the next Section, we will present a crossover from the asymptotic Ising-like critical behavior to the mean-field-like tricritical behavior near the theta point.

First, we consider the consequences of the Flory-Huggins theory in the case of very high degree of polymerization. In this case, the critical composition becomes a small physical parameter, and the appropriate expansion of the free energy could be given through an expansion of the osmotic pressure. The osmotic pressure, $\Pi$, is expressed through  the difference of the chemical potential of the pure solvent, $\hat{\mu}_\text{s}^0$, and its value in solution, $\hat{\mu}_\text{s}$, as $\Pi =\rho(\hat{\mu}_\text{s}^0 - \hat{\mu}_\text{s})$, where $\rho$ is the density. The osmotic pressure (expressed in a reduced form as $\hat{\Pi} = \Pi/\rho k_\text{B}T$) is related to the Gibbs energy through a Legendre transform, $\hat{\Pi} = \phi\hat{\mu}_\text{ms}-\hat{G}$, and may be represented as a virial expansion around the theta point as\cite{Rubinstein_Polymer_2003,Withers_Virial_2003}
\begin{equation}\label{Eq_OsmoticExpans}
	\hat{\Pi}  \approx \frac{\phi}{N} + B\phi^2 + C\phi^3\text{,}
\end{equation}
where the second-order virial coefficient is related to the distance to the theta temperature by $B = (1/2)(T-\Theta)/T$, while the third-order coefficient is $C=1/3$. 

In the scaling theory, the osmotic pressure, like ordinary pressure, is a field-dependent free energy. This is in contrast to the Gibbs energy, which is a ``density''-dependent free energy. One may interconvert between field-dependent and density-dependent free energies through Legendre transforms. For instance, in single-component substances, the density of the grand canonical free energy, $\Omega$, is related to both the pressure, $P$, of the system and the Helmholtz free energy, $A$, by a Legendre transform of the form $\Omega/V = - P = \rho A-\rho\mu$, where $\rho$ is the density and the field variable is the chemical potential, $\mu$.

To describe the theta-point thermodynamics in terms of an appropriate density-dependent free energy, $\hat{\Phi}$, we first need to introduce an appropriate order parameter for such an energy. De Gennes \cite{deGennes_Collapse_1975,deGennes_Book_1979}, based on the ideas of Edwards \cite{Edwards_Polymers_1965}, defined the order parameter characterizing the long polymer chain as a vector, ${\psi}(\vec{r})$, associated with the statistical behavior of the polymer-chain ends \cite{Cloizeaux_Lagragian_1975,Moore_SemiDilute_1977,Schafer_Polymer_1977}. De Gennes proved that by integrating over the length of the polymer chain, $\psi(\vec{r})$ is related to the volume fraction by $|\psi|^2 = \phi$ \cite{deGennes_Book_1979}. Assuming that the polymer-chain free energy, $\hat{\Phi}$, may be expanded in powers of the polymer-chain order parameter, we write the expansion truncated at the sixth-order term as
\begin{equation}\label{Eq_Gen_PolyChain_Free}
	\hat{\Phi} = \frac{1}{2}h_2|\psi|^2 + \frac{1}{4}u_\psi|\psi|^4 + \frac{1}{6}g_\psi|\psi|^6\text{,}
\end{equation}
where the coefficient $g_\psi$ is independent of $T$ and $\phi$, while $u_\psi$ vanishes at the theta point, as shown below. From Eq.~(\ref{Eq_Gen_PolyChain_Free}), we  define the ``ordering'' field $h_1(\vec{r})$, conjugate to the order parameter, as $h_1(\vec{r}) = \partial\hat{\Phi}/\partial[\psi(\vec{r})]$, which, like the order parameter, is also a vector. The second field, $h_2$, conjugate to $|\psi|^2=\phi$, is expected to play a similar role as the reduced distance to the critical temperature, $\Delta\hat{T}$, in Eq.~(\ref{Eq_LndExp_FreeEng}).

To connect the expansion coefficients ($u_\psi$ and $g_\psi$) with physical properties of the polymer chain (\textit{e.g.}, $\phi$, $T$, and $N$), we must relate the polymer-chain free energy, $\hat{\Phi}$, to the osmotic pressure, $\hat{\Pi}$. This can be done through a Legendre transform, $\hat{\Pi}(T,{h}_1) = \psi(\vec{r})\cdot h_1(\vec{r}) - \hat{\Phi}(T,|\psi|)$ - analogous to the Legendre transform between the pressure and the Helmholtz free-energy density. Performing this Legendre transform, with $\hat{\Phi}$ given by Eq.~(\ref{Eq_Gen_PolyChain_Free}), we find
\begin{equation}\label{Eq_Gen_PolyChain_Osmot}
	\hat{\Pi} = \frac{1}{2}h_2|\psi|^2 + \frac{3}{4}u_\psi|\psi|^4 + \frac{5}{6}g_\psi|\psi|^6\text{.}
\end{equation}
By definition, Eq.~(\ref{Eq_Gen_PolyChain_Osmot}) is the reduced osmotic pressure, such that the coefficients of the Landau expansion of the polymer-chain free energy in Eq.~(\ref{Eq_Gen_PolyChain_Osmot}) must be related to the coefficients of the virial expansion in Eq.~(\ref{Eq_OsmoticExpans}). Substituting $\phi = |\psi|^2$ in Eq.~(\ref{Eq_OsmoticExpans}) and matching the corresponding terms, we obtain the following relationships. From the coefficients of the $|\psi|^2$ term,
\begin{equation}\label{Eq_LndCoeffs_2nd}
	 h_2 = \frac{2}{N} - 4B|\psi|^2 - 3C|\psi|^4\text{,}
\end{equation}
while from the coefficients of the $|\psi|^4$ and $|\psi|^6$ terms
\begin{equation}\label{Eq_LndCoeffs_4thn6th}
	u_\psi = 4B \quad\quad \text{and} \quad\quad g_\psi = 3C\text{.}
\end{equation}
With use of Eqs.~(\ref{Eq_LndCoeffs_2nd}) and (\ref{Eq_LndCoeffs_4thn6th}), the expansion of the polymer-chain free energy can be expressed through the virial expansion coefficients as
\begin{equation}\label{Eq_PolyFreeEng_Final}
	\hat{\Phi} = -\frac{1}{N}|\psi|^2 + B |\psi|^4 + C|\psi|^6\text{.}
\end{equation}
This expression describes the behavior of the polymer solution in the close vicinity of the theta point.

\begin{figure}[t]
	\centering
	\includegraphics[width=0.8\textwidth]{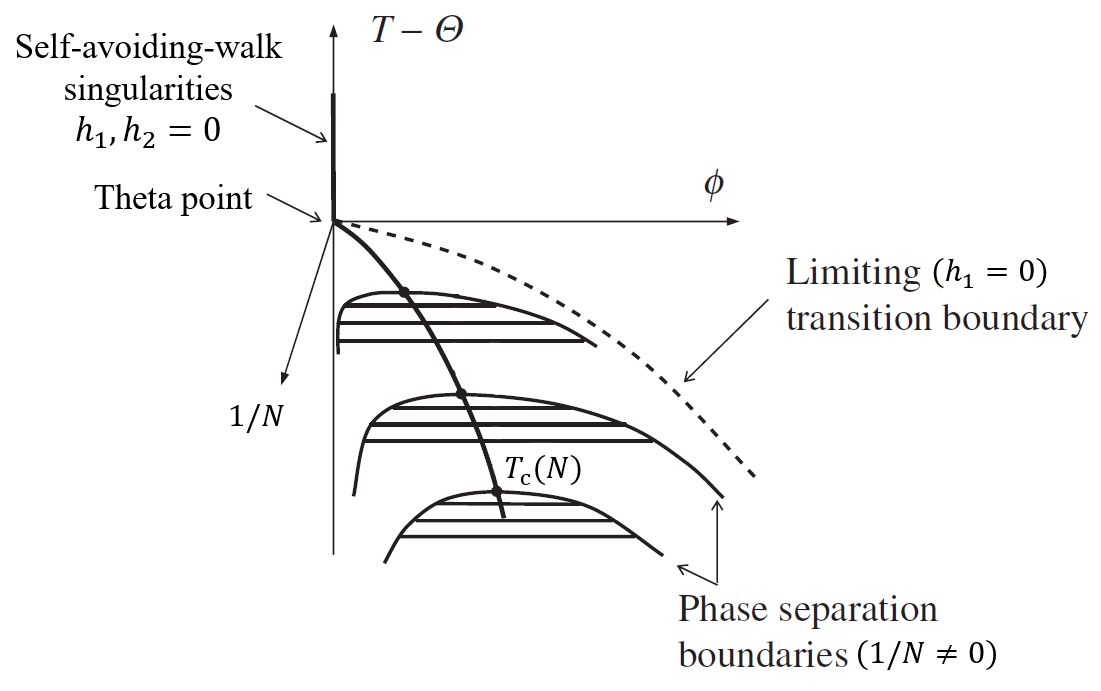}
	\caption{Polymer-chain scaling fields and phase separation of polymer solution in the vicinity of the theta-point (tricritical point). For temperatures $T>\Theta$, the self-avoiding-walk singularity line (``lambda-line'') for a polymer solution in the limit  $N\to\infty$ is analogous to a line of second-order phase transitions ($h_1=h_2=0$). For temperatures $T<\Theta$, the lambda-line transforms into a first-order phase transition line. The line splits into two branches: the polymer branch (dashed curve) and pure solvent (vertical line), where the critical line of demixing at finite $N$, is shown by the solid black curve.}
	\label{Fig_ThetaPhaseBehavior}
\end{figure}

One can observe that the polymer-chain free energy, Eq.~(\ref{Eq_PolyFreeEng_Final}), contains all of the features present in the Landau theory of tricritical phase transitions \cite{LL_Stat_Phys}. Phenomenologically, a tricritical point separates lines of second-order phase transitions from first-order phase transitions, as shown in Fig.~\ref{Fig_Tricrit_Comp}. In the Landau theory, a tricritical point exists where both the coefficients of the second- and fourth-order terms in Eq.~(\ref{Eq_PolyFreeEng_Final}) are zero. The renormalization of the fourth-order term may happen due to the coupling between order parameters (a scalar, $\varphi$, and a vector, $\psi$), described by a coupling term in the Landau expansion, in the form of $\lambda \varphi |\psi|^2$ \cite{Anisimov_Coupling_1981,Knobler_PT_1984,anisimov_critical_1991}. This coupling term originally exists in the Flory-Huggins Gibbs energy, Eq.~(\ref{Eq_FloryFreeEng}), as $\varpi\phi^2 = \varpi\phi|\psi|^2$. Therefore, the Flory-Huggins interaction parameter, $\varpi$, plays the same role as the coupling constant $\lambda$ in the theory of tricriticality. The renormalization of the fourth-order term in the polymer-chain free energy, Eq.~(\ref{Eq_PolyFreeEng_Final}), occurs upon the approach to the theta temperature as $B =1/2-\varpi = 1/2(1-\Theta/T)$, in accordance with the Flory-Huggins theory.

As shown in Eq.~(\ref{Eq_PolyFreeEng_Final}), this occurs in polymer solutions when $N\to\infty$ and $T\to\Theta$, which indicates that the theta point is a tricritical point. In polymer solutions, states above the theta temperature, along the line of zero polymer volume fraction $\phi=0$, correspond to the ``critical states,'' characterized as self-avoiding-walk singularities \cite{deGennes_Book_1979,Fisher_Coulombic_1994}. While no real second-order phase transition occurs for these states, the line (along $\phi=0$ with $N\to\infty$) is analogous to the critical (\textit{lambda}) line in other systems with tricritical phase transitions, such as in mixtures of He$^3$-He$^4$ \cite{Vollhardt_He_1990,Schmitt_He_2015}. For temperatures below the theta temperature, the system is phase separated, in which the polymer chain entirely collapses into one phase. This collapse of the polymer chain is accompanied by the emergence of a finite polymer volume fraction in this phase, and thus, can be regarded as a first-order phase transition. In Eq.~(\ref{Eq_PolyFreeEng_Final}), this behavior is observed depending on the sign of the coefficient of the fourth-order term, $B$. When $B>0$, there is no phase separation, while when $B<0$ the phase separation is first-order. 

The value for the polymer-chain order parameter, $\psi_0(\vec{r})$, which lies along coexistence for $T<\Theta$ and along the critical composition for $T>\Theta$, is given from the minimization of the polymer-chain free energy, $\partial\hat{\Phi}/\partial[\psi(\vec{r})] = 0$, as
\begin{equation}
	|\psi_0|^2 = 
	\begin{cases}
		0 & \text{for $T>\Theta$} \\
		-\dfrac{B}{2C} \pm \sqrt{\left(\dfrac{B}{2C}\right)^2 + \dfrac{1}{N}} & \text{for $T\le \Theta$} 
	\end{cases}\text{.}
\end{equation}
Thus, the ``angle-like'' behavior of the coexistence curve is described in the limit $N\to\infty$ (for $T\le \Theta$) as
\begin{equation}\label{Eq_CoexOrderParam}
	|\psi_0(N\to\infty)|^2 = \varphi_0 = \frac{3}{2}\left(\frac{\Theta - T}{T}\right)\text{\red{.}}
\end{equation}
We note that the same result for the phase behavior is predicted from Flory-Huggins theory, by taking the excess chemical potential of the solvent to be zero in the osmotic pressure expansion, Eq.~(\ref{Eq_OsmoticExpans}), when $N\to\infty$. Similarly, in the limit $T\to\Theta$, the spontaneous polymer-chain order parameter is given by
\begin{equation}
	|{\psi}_0(T\to\Theta)|^2 = \varphi_0 = \sqrt{1/N}\text{.}
\end{equation}
The fact that when $B = 0$, the spontaneous value of the order parameter scales with the leading coefficient to the power $1/4$, is another feature common to tricritical phase transitions \cite{LL_Stat_Phys}.

When the degree of polymerization is large, but still finite, the behavior of the polymer chain is affected by the degree of polymerization. From the polymer-chain free energy, we can define the ordering-field for the polymer-chain, $h_1(\vec{r})$, as
\begin{equation}\label{Eq_PolyOrderField}
	h_1(\vec{r}) = \frac{\partial \hat{\Phi}}{\partial [\psi(\vec{r})]}\bigg|_{\red{h_2}} = \red{\psi(\vec{r})\left[h_2 + 4B|\psi|^2 + 3C|\psi|^4\right] } = \frac{2}{N}\psi(\vec{r})\text{.}
\end{equation}
Physically, the ordering-field in the polymer chain can be related to the energy difference between the chain ends. The closer the polymer chain ends are to each other, the more the chain is affected by this ordering field.

The response of the polymer-chain to the ordering field is given by the polymer-chain susceptibility defined as, $\hat{\chi}_\psi = (\partial^2\hat{\Phi}/\partial[ \psi(\vec{r})]^2)^{-1}$. Along the spontaneous value of the order parameter, $|\psi_0|$, and in the limit $N\gg B$, the inverse susceptibility is given by
\begin{equation}
	\hat{\chi}_\psi^{-1} \propto \frac{1}{N} + B^2\text{.}
\end{equation}
The susceptibility scales with the degree of polymerization at the theta temperature, as $\hat{\chi}_\psi (T\to\Theta)\propto N$, while in limit $N\to\infty$, the susceptibility is inversely proportional to the distance to the theta point, as $\hat{\chi}_\psi (N\to\infty) \propto B^{-2}$. Moreover, the behavior of the correlation length of the polymer-chain order parameter, $\xi_\psi$, in the vicinity of the theta point can be determined through a Landau Ginzburg functional, similar to Eq.~(\ref{Eq_FreeFunc}), such that 
\begin{equation}\label{Eq_Ch5_Corr_PolyChain}
	\xi_\psi = \sqrt{c_\psi\hat{\chi}_\psi} \propto \frac{ \sqrt{N}}{1+\sqrt{N}B^2}\text{,}
\end{equation}
where it is important to note that $c_\psi$ only depends on the size of the monomer. Equation~(\ref{Eq_Ch5_Corr_PolyChain}) has two key limits: at the theta temperature $\xi_\psi (T\to\Theta)\propto \sqrt{N}$ and for an infinite degree of polymerization, $\xi_\psi (N\to\infty) \propto B^{-1}\sim (\Theta - T)^{-1}$. 

A remarkable feature of tricriticality is the fact that in three dimensions, the system near tricritical points demonstrates essentially mean-field behavior. For Ising-like systems, the marginal dimension that separates mean-field criticality and scaling criticality is $d=4$. For tricritical systems, the marginal dimension is $d=3$ \cite{Lawrie_PT_1984,Hager_Crossover_2002,Kostko_Crossover_2005}. Therefore, the crossover between criticality and theta-point tricriticality is essentially a crossover between asymptotic Ising-like critical behavior and mean-field tricritical behavior, which is discussed in detail in the next Section.

\section{Crossover between \red{Ising} Criticality and Theta-Point Tricriticality \label{Sec3_Crossover}} 

\begin{figure}[b]
	\centering
	\includegraphics[width=0.7\textwidth]{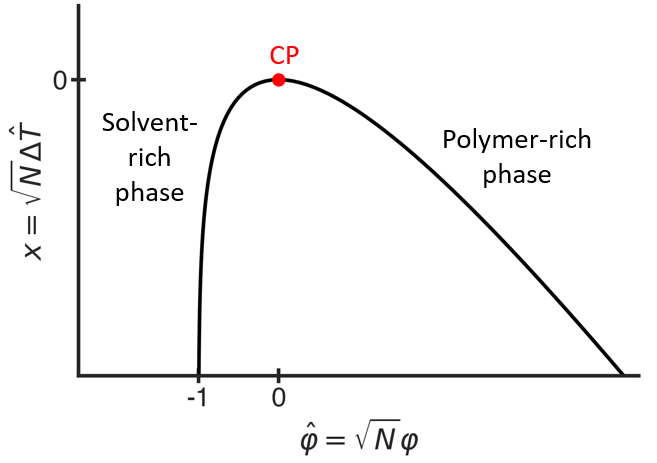}
	\caption{A universal phase diagram in Flory-Huggins theory, as suggested by Widom \cite{Widom_Scaling_1993}, is achieved by rescaling the volume-fraction order parameter, $\varphi$, and the reduced distance to the critical temperature, $\Delta\hat{T}$, by the square root of the degree of polymerization.}
	\label{Fig_Widom_Crossover}
\end{figure}

As first demonstrated by Widom \cite{Widom_Scaling_1993}, the phase coexistence in the mean-field Flory-Huggins theory may be represented in a universal way by rescaling the order parameter, $\varphi=\phi-\phi_\text{c}$, and the reduced distance to the critical temperature, $\Delta\hat{T} = (T-T_\text{c})/T_\text{c}$, by $\sqrt{N}$. A schematic of the universal mean-field-scaled behavior is illustrated in Fig.~\ref{Fig_Widom_Crossover}. Following Widom \cite{Widom_Scaling_1993}, we define a ``Widom variable'' as $x = \sqrt{N}|\Delta\hat{T}|$, and since the critical volume fraction scales as $\phi_\text{c}\approx 1/\sqrt{N}$ for large $N$, we define the rescaled order parameter as $\hat{\varphi} = \varphi/\phi_\text{c} \approx \sqrt{N}\varphi$.

Widom demonstrated that there are two key limits of such scaled behavior \cite{Widom_Scaling_1993}. First, asymptotically close to the critical point, where $\Delta\hat{T}\gg \sqrt{N}$ and $x\ll 1$, the order parameter along coexistence is $\hat{\varphi}= \sqrt{3x}$. This result also corresponds to the behavior of the Landau expansion of the Flory-Huggins theory. However, Widom's approach also elucidates the behavior of the polymer solution in the alternative limit, where $\Delta\hat{T}\ll \sqrt{N}$ and $x\gg 1$. In this limit, the order parameter along coexistence adopts two solutions: for the polymer-rich phase, $\hat{\varphi} = (3/2)x$, while for the solvent-rich phase, $\hat{\varphi} = -1$, which corresponds to the ``angle-like'' shape of the coexistence observed in the theta-point (tricritical) limit, Eq.~(\ref{Eq_CoexOrderParam}).

As already discussed, the actual (fluctuation-induced) behavior of the phase coexistence asymptotically close to the critical point follows the Ising-model universality class. Thus, in the region where $x\ll 1$, according to the scaling theory, $\hat{\varphi}\simeq x^\beta$, where $\beta = 0.326$ (see Table~\ref{Table_CritExp}). The region where the mean-field approximation becomes insufficient, and the effects of critical fluctuations must be taken into account, is given by the ``Ginzburg criterion'' \cite{Anisimov_Crossover_Gen_1992,Anisimov_CriticalRegion_2000}, $|\Delta\hat{T}|\ll |\Delta\hat{T}|_\times = N_\text{G}$, where $N_\text{G}$ is the Ginzburg number and $|\Delta\hat{T}|_\times = (T_\times-T_\text{c})/T_\text{c}$. The Ginzburg number can be estimated by the following expression \cite{Anisimov_Crossover_Gen_1992,Anisimov_CriticalRegion_2000}
\begin{equation}\label{Eq_GinzNumb}
	|\Delta\hat{T}|_\times \approx u_0^2\left(\frac{\upsilon_0}{\bar{\xi}_0^3}\right)^2\text{,}
\end{equation}
where $u_0$ is the fourth-order coefficient in the Landau expansion and $\upsilon_0^{1/3}$ is of the order of the range of interactions, $\xi_0$. The Ginzburg number indicates the crossover temperature from mean-field behavior away from the critical point to Ising-like behavior in the asymptotic vicinity of the critical point. For ordinary second-order phase transitions, the Ginzburg number is controlled by the range of interactions, $\xi_0$, while $u_0$ and $\upsilon_0$ are effectively constant \cite{Kim_Crossover_2003}. For polymer solutions, $u_0 = 2\sqrt{N}$, as given in Eq.~(\ref{Eq_LndExp_FreeEng}), $\upsilon_0$ is the volume of the monomer, and $\bar{\xi}_0\approx \upsilon_0^{1/3} N^{1/4}$ is the amplitude of the correlation length in the mean-field approximation. Thus, Eq.~(\ref{Eq_GinzNumb}) can be simplified as $|\Delta\hat{T}|_\times = c/\sqrt{N}$, where $c$ is a system dependent constant of the order of unity. Analogously,  at $x =\sqrt{N}|\Delta\hat{T}|=1$, the Widom crossover indicates the crossover temperature from mean-field tricritical behavior in the vicinity of the theta point to the behavior (either mean-field or Ising-like) in the vicinity of the critical-demixing point. 

The Widom procedure for the crossover between the mean-field critical regime and the theta-point tricritical regime in polymer solutions was generalized by Anisimov and co-workers to include the effects of critical fluctuations in the vicinity of the critical-demixing point \cite{Povodyrev_Crossover_1999,Agayan_Crossover_2001,Hager_Crossover_2002,Kostko_Crossover_2002,Kostko_Crossover_2005,Anisimov_PolyTricriticality_2005}. We refer to this approach as the ``generalized Widom crossover,'' and in this review, we present a simplified version of \red{this} crossover, emphasizing the essential physics of this phenomenon. 

\begin{figure}[t]
	\centering
	\includegraphics[width=0.7\textwidth]{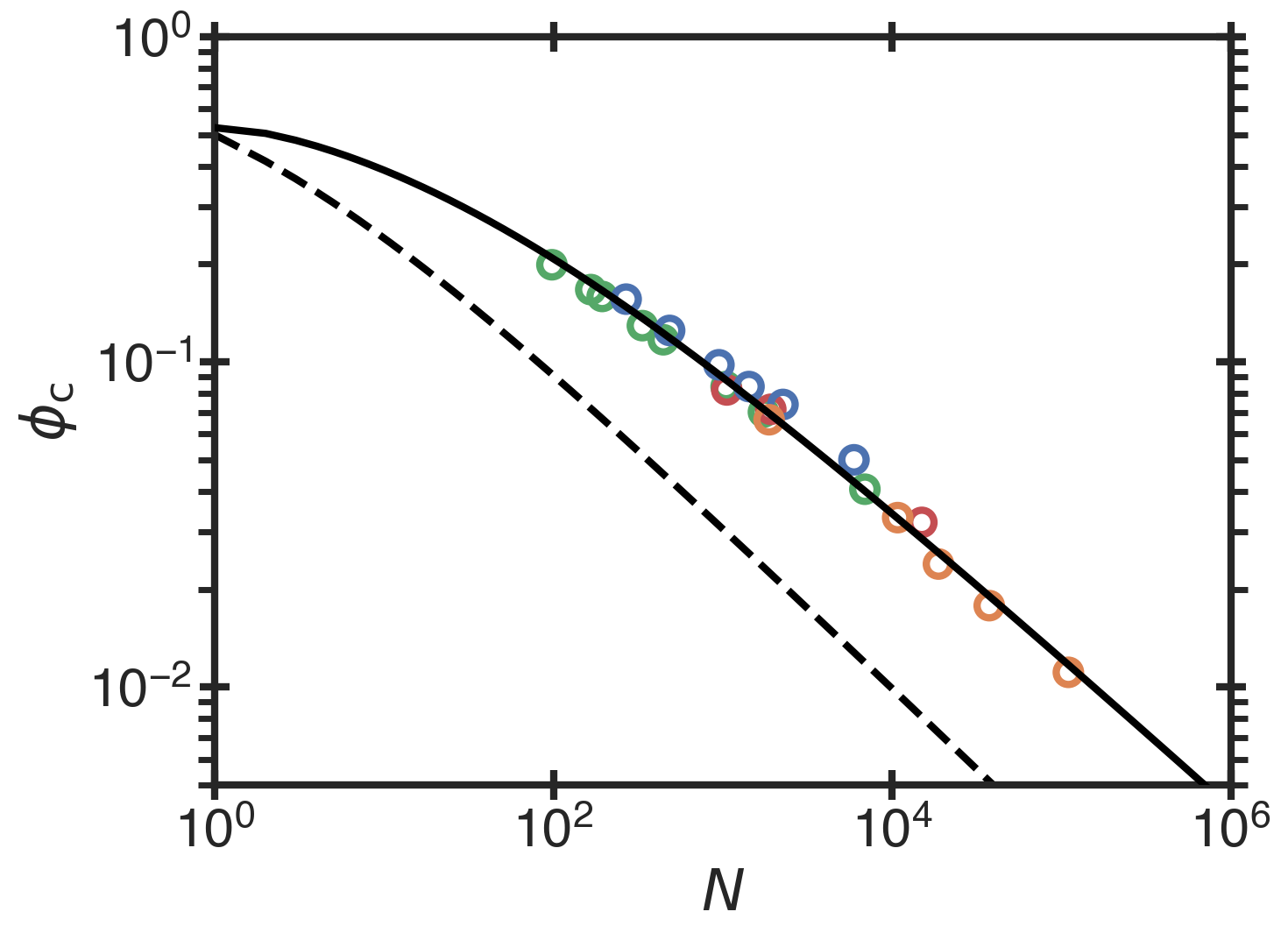}
	\caption{Critical volume fraction, $\phi_\text{c}$ of solutions of polystyrene in cyclohexane measured as a function of the degree of polymerization, $N$, by Kojima and Nakata \textit{et al.}\cite{Kojima_Coex_1975,Nakata_Coex2_1975,Nakata_Coex1_1978} (red), polystyrene in cyclohexane by Anisimov \textit{et al.}\cite{Kostko_Crossover_2005} (orange), polystyrene in methylcyclohexane by Dobashi \textit{et al.}\cite{Dobashi_MethPS_1980} (green), and PMMA in 3-octanone by Xia \textit{et al.}\cite{Xia_PMMA_1992,Xia_PMMA_1996} (blue). The solid curve represents Eq.~(\ref{Eq_LogCorrect_Crit_Params}), with coefficient $a = 0.95$, while the dashed curve represents the prediction from the Flory-Huggins theory, Eq.~(\ref{Eq_FloryCritParams}).}
	\label{Fig_PhicNdepend}
\end{figure}

First, in the vicinity of the theta point, there are logarithmic corrections to the mean-field behavior, induced by fluctuations, since the marginal dimension for tricriticality is $d=3$, that must be accounted for \cite{Lawrie_PT_1984,Hager_Crossover_2002,Kostko_Crossover_2005}. In this case, the classical chain-length dependence of the critical parameters, $\phi_\text{c}$ and $T_\text{c}$, as given by Eq.~(\ref{Eq_FloryCritParams}), are modified by tricritical fluctuations. The modified critical parameters are\cite{Hager_Crossover_2002}
\begin{equation}\label{Eq_LogCorrect_Crit_Params}
	\phi_\text{c} = \frac{1}{1 + a\sqrt{N}(1 + \ln N)^{-1/2}} \quad\quad \text{and} \quad\quad \frac{\Theta-T_\text{c}}{\Theta}\approx \frac{1}{(\ln N)^{3/22}\sqrt{N}}\text{,}
\end{equation}
where the coefficient $a$ is on the order of unity. Note that there is a significant effect of logarithmic fluctuations on the critical volume fraction, as shown in Fig.~\ref{Fig_PhicNdepend}, as compared with the prediction of the Flory-Huggins theory, Eq.~(\ref{Eq_FloryCritParams}). Meanwhile, the effect of logarithmic corrections on the critical temperature is relatively small, but still consequential.

In the generalized Widom crossover, as a result of the logarithmic correction to the theta-point tricritical behavior, a more accurate representation for the Widom variable, $x$, is $\tilde{x} = (T-T_\text{c})/(\Theta -T_\text{c})$, where the critical temperature is measured experimentally and the theta temperature is obtained from extrapolation \cite{Hager_Crossover_2002,Kostko_Crossover_2002,Kostko_Crossover_2005,Anisimov_PolyTricriticality_2005}. In the Flory-Huggins theory, $\tilde{x} = x$, as $(\Theta-T_\text{c})/T_\text{c} \approx 1/\sqrt{N} = \phi_\text{c}$ for large $N$. Moreover, the reduced order parameter, $\hat{\varphi} = (\phi-\phi_\text{c})/\phi_\text{c}$, is rescaled with a critical volume fraction determined from experiment. From here, a simple interpolation function, with a crossover parameter given through the Widom variable, is utilized to modify the universal variables ($\tilde{x}$ and $\hat{\varphi}$) to match the scaling limits in the asymptotic Ising-like critical regime. The crossover parameter responsible for the transition from Ising-like criticality to theta-point tricriticality is given as the ratio $z = \xi/R_\text{g}$ \cite{Povodyrev_Crossover_1999,Agayan_Cross_2000,Hager_Crossover_2002,Kostko_Crossover_2002,Kostko_Crossover_2005,Anisimov_PolyTricriticality_2005}. We note that in the Flory-Huggins theory, the crossover variable \red{reduces} to the Widom variable as $z = \tilde{x}^{-1/2}$. Therefore, the crossover behavior of the phase coexistence in polymer solutions can be represented by a simple generalization of the Widom crossover as
\begin{equation}\label{Eq_Cross_Cxc}
	\hat{\varphi} \approx \hat{\varphi}_\text{MF}\left(1+z^{2}\right)^{(1-2\beta)/4\nu}\text{,}
\end{equation}
where the amplitude function, $\hat{\varphi}_\text{MF}$, is predicted from the mean-field Flory-Huggins-like theory. The Ising critical exponent is $\beta =0.326$ (see Table~\ref{Table_CritExp}). \red{Equation}~(\ref{Eq_Cross_Cxc}) has the limits: $\hat{\varphi}(z\ll 1) \to \hat{\varphi}_\text{MF}$ and $\hat{\varphi}(z\gg 1) \to B_0|\Delta\hat{T}|^\beta$, where $B_0\sim N^{-0.337}$ (see Table~\ref{Table_deGenSanchez}).

\begin{figure}[t]
	\centering
	\includegraphics[width=0.49\textwidth]{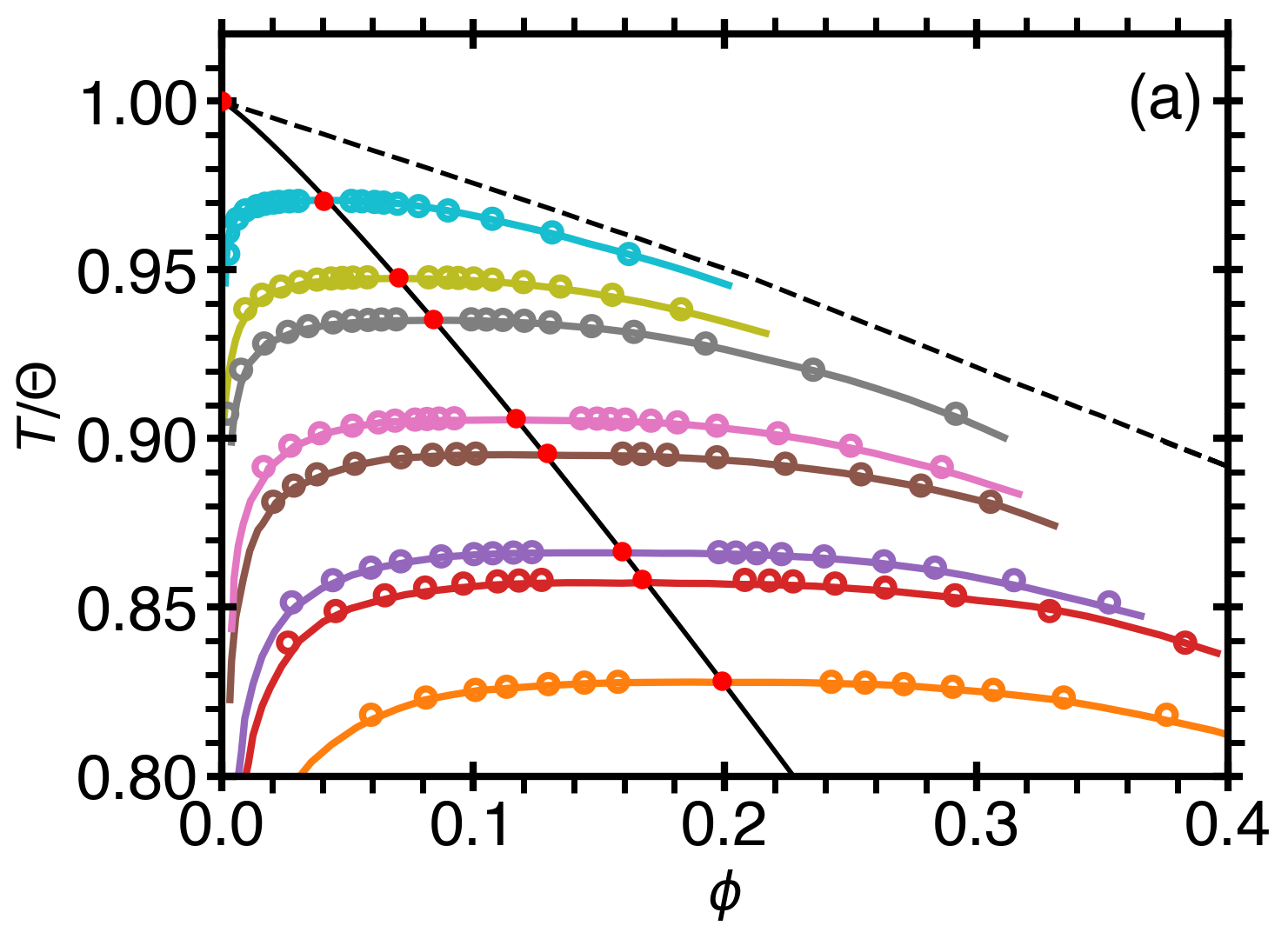}
	\includegraphics[width=0.49\textwidth]{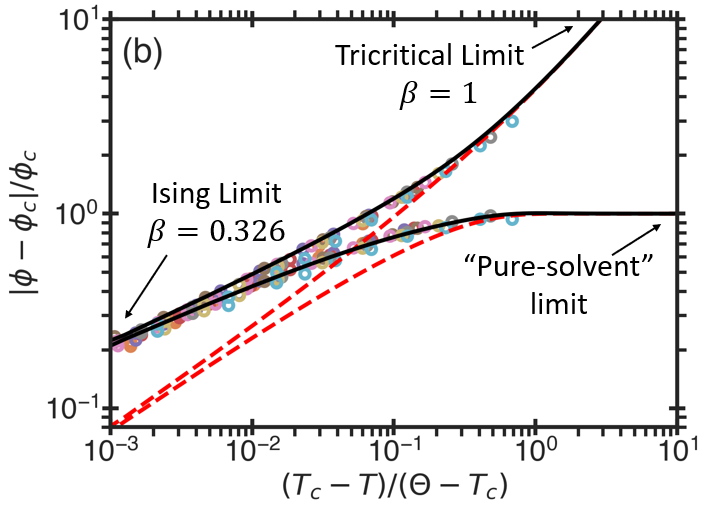}
	\caption{Phase diagram and universal behavior of a solution of polystyrene in methylcyclohexane. (a) Phase behavior of the polymer solution. The open circles are experimental data\cite{Dobashi_MethPS_1980}, while the curves are the theoretical prediction \cite{Agayan_Cross_2000}. (b) Rescaling the volume fraction and temperature generates a universal phase diagram. As shown\red{,} the polymer-rich phase exhibits a crossover from the asymptotic Ising-like critical region to the tricritical theta-point region, while the solvent-rich phase exhibits a crossover from the asymptotic Ising-like region to the limit of pure solvent.}
	\label{Fig_Scaling_Crossover}
\end{figure}

In a similar form to the phase coexistence behavior Eq.~(\ref{Eq_Cross_Cxc}), the crossover expressions for the susceptibility and correlation length are given by\cite{Kostko_Crossover_2005,Anisimov_PolyTricriticality_2005}
\begin{align}
	\hat{\chi}^{-1} &\approx \hat{\chi}^{-1}_\text{MF} (1+z^2)^{(\gamma-1)/2\nu} 
	\label{Eq_Cross_Sus}\text{\red{,}}\\
	\xi &\approx \xi_\text{MF}(1+z^2)^{(2\nu-1)/4\nu}\label{Eq_Cross_Corr}\text{\red{,}}
\end{align}
where the amplitude functions are the susceptibility and correlation length predicted by the mean-field Flory-Huggins-like theory, $\hat{\chi}^{-1}_\text{MF} = \bar{\Gamma}_0|\Delta\hat{T}|$ and $\xi_\text{MF} = \bar{\xi}_0|\Delta\hat{T}|^{-0.5}$ (where $\bar{\Gamma}_0\sim N^0$ and $\bar{\xi}_0\sim N^{-0.25}$ are the mean-field amplitudes), see Eq.~(\ref{Eq_FreeFunc}). Asymptotically close to the critical point (for $z\gg 1$), Eqs.~(\ref{Eq_Cross_Sus}) and (\ref{Eq_Cross_Corr}) satisfy the critical power laws, $\hat{\chi}=\Gamma_0|\Delta\hat{T}|^{-\gamma}$ and $\xi = \xi_0|\Delta\hat{T}|^{-\nu}$ (see Table~\ref{Table_CritExp}). For example, consider the phase behavior of the eight different solutions of polystyrene in methylcyclohexane with different degrees of polymerization\cite{Dobashi_MethPS_1980,Hager_Crossover_2002} presented in Fig.~\ref{Fig_Scaling_Crossover}a. The universal crossover behavior of this system is shown in Fig.~\ref{Fig_Scaling_Crossover}b.

\begin{figure}[t]
	\centering
	\includegraphics[width=0.49\textwidth]{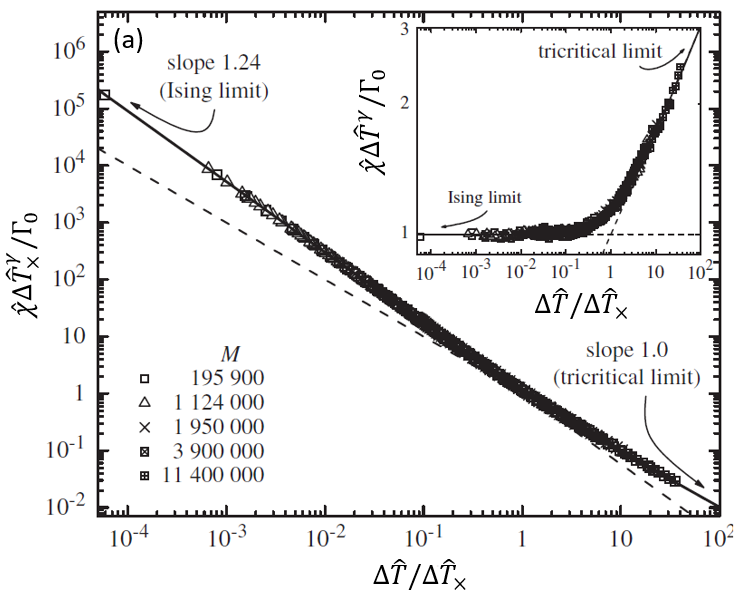}
	\includegraphics[width=0.49\textwidth]{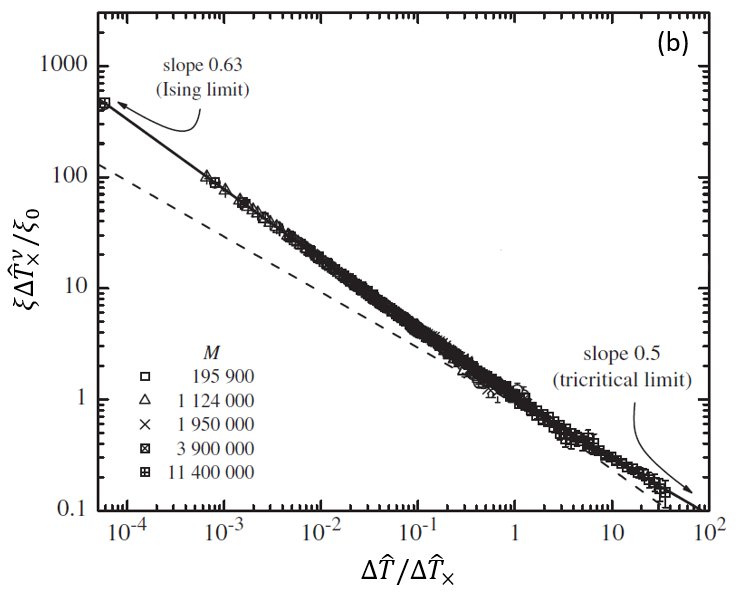}
	\caption{Universal scaling behavior of thermodynamic properties. (a) Scaled susceptibility and deviation of the susceptibility from Ising critical behavior (shown as inset) as a function of the normalized distance from the critical temperature,  $\Delta\hat{T}/\Delta\hat{T}_\times$, for polystyrene–cyclohexane solutions with five	different molecular weights of polystyrene. (b) The scaled correlation length for the same data. The symbols indicate experimental data, while the solid curve represents the crossover theory. The dashed lines represent the two limiting behaviors: Ising asymptotic behavior and theta-point (tricritical) behavior. Adapted with permission from \textit{Phys. Rev. E}\cite{Kostko_Crossover_2005}.}
	\label{Fig_SusCorr_Crossover}
\end{figure}

By rescaling the thermodynamic properties, we may represent the crossover phenomenon in a universal form as
\begin{align}
	\tilde{\varphi} &\equiv \frac{\hat{\varphi}|\Delta\hat{T}|_\times^{-\beta}}{B_0} \approx \frac{(1+z^{2})^{(1-2\beta)/4\nu}}{(\Delta\hat{T}_\times/\Delta\hat{T})^{1/2}}\text{,}\\
	\tilde{\chi} & \equiv \frac{\chi|\Delta\hat{T}|_\times^\gamma}{\Gamma_0} \approx \frac{(1+z^2)^{(1-\gamma)/2\nu}}{\Delta\hat{T}/\Delta\hat{T}_\times}\text{,}\\
	\tilde{\xi} &\equiv \frac{\xi|\Delta\hat{T}|_\times^\nu}{\xi_0} \approx \frac{(1+z^2)^{(2\nu-1)/4\nu}} {(\Delta\hat{T}/\Delta\hat{T}_\times)^{1/2}}\text{.}
\end{align}
Figure~\ref{Fig_SusCorr_Crossover} \red{shows} the universal susceptibility (a) and correlation length (b) for five different degrees of polymerization in a polystyrene-cyclohexane solution \cite{Kojima_Coex_1975,Nakata_Coex2_1975,Nakata_Coex1_1978,Dobashi_MethPS_1980,Xia_PMMA_1992,Xia_PMMA_1996,Kostko_Crossover_2005}. The inset of Fig.~\ref{Fig_SusCorr_Crossover}a shows the deviation of the susceptibility from the asymptotic Ising-like behavior in the vicinity of the \red{theta}-point. In their rescaled forms, the thermodynamic properties exhibit the middle of their crossover at the Widom crossover point, $x = \Delta\hat{T}/\Delta\hat{T}_\times = 1$.

Another phenomenon associated with the crossover to theta-point tricriticality is the dynamic coupling of the diffusive relaxation of the critical concentration fluctuations discussed in Sec.~\ref{Sec2_CritPhenom}, and entanglements/disentanglements of long polymer chains (commonly referred in the literature as a ``viscoelastic mode''\cite{Tanaka_Dynamic_2002}). Experimental verification of the dynamic coupling between these two modes was demonstrated independently by Tanaka \textit{et al.} \cite{Tanaka_Dynamic_2002} and Kostko \textit{et al.} \cite{Kostko_Dynamics_2002,Kostko_Crossover_2005}. According to the theory of de Gennes and Brochard \cite{Brochard_Dynamics_1977,Brochard_Dynamics2_1983}, the normalized time-dependent intensity correlation function, $g_2(t)$, is given in the form
\begin{equation}\label{Eq_CorrelationFunction_Dynamics}
	g_2(t)-1 = \left[f_+e^{-\omega_+t} + f_-e^{-\omega_-t}\right]^2\text{,}
\end{equation}
with decay frequencies given by
\begin{equation}\label{Eq_DynamicModes}
	\omega_\pm = \frac{\omega_\psi}{2}\left[1 + q^2\xi_\psi^2 + \frac{\omega_q}{\omega_\psi} \pm \sqrt{\left(1 + q^2\xi_\psi^2 + \frac{\omega_q}{\omega_\psi}\right)^2 - 4\frac{\omega_q}{\omega_\psi}} \right]\text{,}
\end{equation}
and corresponding amplitudes $f_+$ and $f_-$ given by
\begin{equation}\label{Eq_DynamicAmps}
	f_\pm = \pm \frac{\omega_\pm -\omega_\psi(1+q^2\xi_\psi^2)}{\omega_+ - \omega_-}\text{,}
\end{equation}
where $\omega_\pm$ are the coupled modes, in which the relaxation of the faster mode is designated as, $\omega_+$, and consequently, the slower mode is designated as $\omega_-$. In previous works, $\xi_\psi$ was defined as a mesoscopic viscoelastic length scale, which we identify with the correlation length of the polymer chain, given by Eq.~(\ref{Eq_Ch5_Corr_PolyChain}). In Eqs.~(\ref{Eq_DynamicModes}) and (\ref{Eq_DynamicAmps}), $\omega_q$ represents the uncoupled $q$-dependent diffusion relaxation mode, as specified through the diffusion equations given by Eqs.~(\ref{Eq_CritDiff}) and (\ref{Eq_DiffBack}), and $\omega_\psi$ is an uncoupled $q$-independent viscoelastic relaxation mode associated with the polymer chain and assumed to be inversely proportional to the viscosity of the solution, which diverges in the limit $N\to\infty$, as $\omega_\psi\propto \eta(T)^{-1}$ \cite{Kostko_Diffusion_2007}. The diffusive relaxation mode, $\omega_q$ is associated with the conserved order parameter, $\varphi = \phi-\phi_\text{c}$, while the viscoelastic mode is associated with the non-conserved polymer-chain order parameter $|\psi|$ \cite{Tanaka_Dynamic_2002,Kostko_Dynamics_2002,Kostko_Diffusion_2007}. The coupling between the two dynamic modes reflects the coupling between these two order parameters that is responsible for the emergence of the theta-point tricriticality.

\begin{figure}[t]
	\centering
	\includegraphics[width=0.7\textwidth]{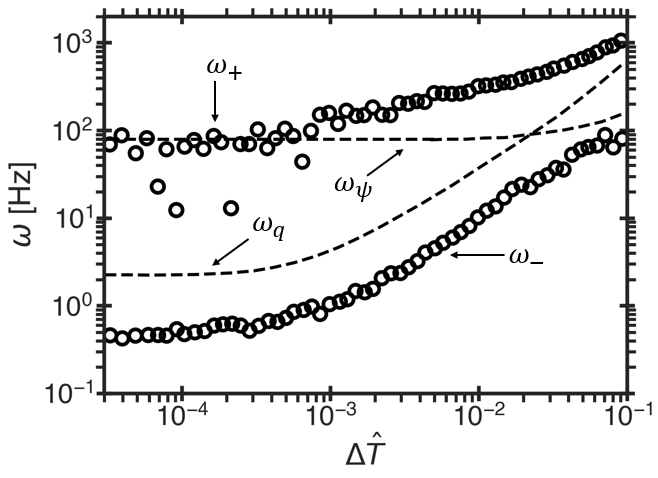}
	\caption{Decay frequencies, $\omega_\pm$, for a solution of polystyrene in cyclohexane, with \red{$N=110,000$}, measured at scattering angle, $\theta =\SI{30}{\degree}$ \cite{Kostko_Diffusion_2007}. The symbols represent the values for the two coupled modes extracted from the fit of the correlation-function data, Eq.~(\ref{Eq_CorrelationFunction_Dynamics}). The points correspond to the faster mode, $\omega_+$ (upper), and the slower mode, $\omega_-$ (lower). The dashed curves represents the theoretical calculations of the uncoupled viscoelastic mode, $\omega_\psi$, and the uncoupled diffusive mode, $\omega_q$.}
	\label{Fig_Avoid_Cross}
\end{figure}

The effect of coupling between the two dynamic modes is most clearly observed in polymer solutions with a high degree of polymerization (closer vicinity to the theta point). For instance, Fig.~\ref{Fig_Avoid_Cross} depicts the dynamic coupling in a solution of polystyrene in cyclohexane with molecular weight of $11.4$ million $\si{\gram/\mole}$ (\red{$N = 110,000$}) \cite{Kostko_Diffusion_2007}. The avoided crossing in the behavior of the fast and slow frequency modes, $\omega_\pm$, is reminiscent of the Landau-Zener effect in the transition dynamics of a two-state quantum system \cite{Landau_Avoid_1932,Zener_Avoid_1932}. As illustrated in Fig.~\ref{Fig_Avoid_Cross}, in the vicinity of the critical point, both the uncoupled viscoelastic and diffusive modes saturate. The diffusive relaxation time saturates at a finite wave number, controlled by the scattering angle, while the viscoelastic relaxation time saturates because the degree of polymerization is finite and $\xi_\psi$ reaches a finite value. In this region, the characteristic frequency of the faster coupled mode scales with the correlation length of the polymer chain as $\omega_+ = \omega_\psi(1+q^2\xi_\psi^2)$, while the frequency of the slower coupled mode scales as $\omega_-=\omega_q(1+q^2\xi_\psi^2)^{-1}$. In the region far away from the critical temperature, the two modes uncouple as $\omega_-\approx \omega_\psi$ and $\omega_+\approx\omega_q$ \cite{Tanaka_Dynamic_2002,Kostko_Dynamics_2002,Kostko_Diffusion_2007}. We conclude that dynamic light-scattering experiments enable one to obtain the polymer-chain correlation length, $\xi_\psi$.

\section{Critical Phenomena in a Polymer Blend}\label{Sec_Polymer_Blend}
Our treatment of the critical behavior in polymer solutions may be generalized to a mixture of two incompressible polymer species, A and B, with degrees of polymerization ${N}_{\text{A}}$ and ${N}_\text{B}$, respectively. Modifying the Flory-Huggins theory to account for the lengths of each species, the Gibbs energy of mixing of the blend, $\hat{G}_\text{bd}$, is given by\cite{Flory_Polymer_1941,Huggins_Solutions_1941,Flory_Theory_1953,Sanchez_Blends_1983,Binder_Blend_1983,Panayiotou_Blends_1989}
\begin{equation}\label{Eq_PolyBlend_FreeEnergy}
	\hat{G}_\text{bd} = \left[\frac{\phi}{{N}_{\text{A}}}\ln\phi + \frac{(1-\phi)}{{N}_\text{B}}\ln(1-\phi)\right] + \varpi\phi(1-\phi)\text{,}
\end{equation}
where the Flory interaction parameter is approximated as $\varpi = \Theta/2T$, see Eq.~(\ref{Eq_FloryFreeEng}). Equation~(\ref{Eq_PolyBlend_FreeEnergy}) has the following two limits: first, in the case when one species is much longer than the other, ${N}_\text{A} \equiv {N} \gg N_\text{B}$ and ${N}_\text{B} \approx 1$, then Eq.~(\ref{Eq_PolyBlend_FreeEnergy}) reduces to the Gibbs energy of mixing in the Flory-Huggins theory, Eq.~(\ref{Eq_FloryFreeEng}). Hereafter, we refer to this case as the asymmetric limit. Second, in the case when each species is of the same size, ${N}_\text{A} = {N}_\text{B} ={N}$, then the contribution from the entropy of mixing, the term in brackets in Eq.~(\ref{Eq_PolyBlend_FreeEnergy}), scales with $1/N$. We refer to this case as the symmetric limit.

The critical volume fraction and critical temperature are determined from the thermodynamic stability conditions, $(\partial^2\hat{G}_\text{bd}/\partial\phi^2)|_{T,N_\text{A},N_\text{B}} = 0$ and $(\partial^3\hat{G}_\text{bd}/\partial\phi^3)|_{T,N_\text{A},N_\text{B}} = 0$, yielding
\begin{equation}\label{Eq_Blend_Crit_Prm}
	\phi_\text{c} = \frac{\sqrt{{N}_\text{B}}}{\sqrt{{N}_\text{A}} + \sqrt{{N}_\text{B}}} \quad\quad \text{and} \quad\quad T_\text{c} = \Theta\left[\frac{1}{\sqrt{{N}_\text{A}}} + \frac{1}{\sqrt{{N}_\text{B}}} \right]^{-2}\text{.}
\end{equation}
In the asymmetric limit, the critical parameters given by Eq.~(\ref{Eq_Blend_Crit_Prm}) reduce to those given by Eq.~(\ref{Eq_FloryCritParams}). In the symmetric limit, $\phi_\text{c} = 1/2$ and $T_\text{c} = \Theta {N}/2$, and, when both polymer chains are sufficiently long, such that ${N}\gg 1$, then $T_\text{c}\to\infty$, and the polymer blend would not mix for any degree of nonideality.

Just as given by Eq.~(\ref{Eq_LndExp_FreeEng}), the truncated Landau expansion may be represented in the form
\begin{equation}\label{Eq_Ch5_BlendLndExpns}
	\Delta \hat{G}_\text{bd} =\frac{1}{2}\tilde{a}_0\Delta\hat{T}\varphi^2 + \frac{1}{4!}\tilde{u}_0\varphi^4\text{,}
\end{equation}
where the expansion coefficients for the blend are given by
\begin{equation}\label{Eq_Blend_a0_u0}
	\tilde{a}_0 = \left(\frac{1}{\sqrt{{N}}_\text{A}}+\frac{1}{\sqrt{{N}}_\text{B}}\right)^2 \quad\quad \text{and}\quad\quad \tilde{u}_0 = \sqrt{{N}_\text{A}}\sqrt{{N}_\text{B}}\left(\frac{1}{\sqrt{{N}}_\text{A}}+\frac{1}{\sqrt{{N}}_\text{B}}\right)^4\text{.}
\end{equation}
The Landau expansion of the Flory-Huggins Gibbs free energy, Eq.~(\ref{Eq_LndExp_FreeEng}), is recovered in the asymmetric limit. In the symmetric limit, the second- and fourth-order coefficients of Eq.~(\ref{Eq_Blend_a0_u0}), are given by $a_0 = 4/{N}$ and $u_0=32/{N}$. In the symmetric blend, any degree of non-ideality will produce phase separation in the limit $N\to\infty$.

The osmotic pressure for the polymer blend is related to the Gibbs energy through a Legendre transform, $\hat{\Pi}_\text{bd}=\phi (\partial\hat{G}_\text{bd}/\partial\phi)|_{T,N_\text{A},N_\text{B}} - \hat{G}_\text{bd}$, and may be represented as a virial expansion around the theta point as\cite{Binder_Blend_1983}
\begin{equation}
	\hat{\Pi}_\text{bd} = \frac{\phi}{{N}_\text{A}} + \frac{1}{2}\left(\frac{1}{{N}_\text{B}} - \frac{\Theta}{T}\right)\phi^2 + \frac{1}{3{N}_\text{B}}\phi^3\text{.}
\end{equation}
Comparing with the asymmetric limit for osmotic pressure, Eq.~(\ref{Eq_OsmoticExpans}), we see that the second and third virial coefficients depend on ${N}_\text{B}$ as
\begin{equation}
	B = \frac{1}{2}\left(\frac{1}{{N}_\text{B}} - \frac{\Theta}{T}\right) \quad\quad \text{and} \quad\quad C = \frac{1}{3{N}_\text{B}}\text{.}
\end{equation}
This dependence on $N_\text{B}$ has important implications for the polymer-chain free energy, Eq.~(\ref{Eq_PolyFreeEng_Final}). Following the derivation in Sec.~\ref{Sec3_Crossover}, the polymer-chain free energy of the blend is given by
\begin{equation}\label{Eq_PolyChainFree_PolyBlend}
	\hat{\Phi}_\text{bd} = -\frac{1}{{N}_\text{A}}|\psi|^2 + \frac{1}{2}\left(\frac{1}{{N}_\text{B}} - \frac{\Theta}{T}\right) |\psi|^4 + \frac{1}{3{N}_\text{B}}|\psi|^6\text{.}
\end{equation}
The condition for theta-point tricriticality is only observed in the asymmetric limit where ${N}_\text{A}\gg {N}_\text{B}\to\infty$. However, in the symmetric limit, the polymer-chain free energy becomes undefined at the theta point as $T\to\Theta$ and ${N}\to\infty$, while $\psi(\vec{r})\to 0$, implying that there is no theta-point tricriticality in a symmetric blend.

It is important to note that, for finite $N$, the polymer blend exhibits Ising-like critical behavior in the vicinity of the liquid-liquid critical point \cite{Meier_BlendCrit_1992,Hair_Blend_1992}, just like all fluid mixtures \cite{fisher_scaling_1983,Anisimov_CriticalRegion_2000}. The size of the critical region is determined through the Ginzburg criterion, Eq.~(\ref{Eq_GinzNumb}), which scales with the degree of polymerization of each species as \cite{Bates_Blend_1990,Hair_Blend_1992,Schwahn_BlendGinz_1995,Wang_Blend_2002}
\begin{equation}\label{Eq_GinzNumb_Blend}
	N_\text{G} \sim \frac{1}{\sqrt{{N}_\text{A}{N}_\text{B}}}\text{.}
\end{equation}
Equation~(\ref{Eq_GinzNumb_Blend}) has the following two limits: in the asymmetric limit, $N_\text{G}\sim 1/\sqrt{{N}}$, just as discussed in Sec.~\ref{Sec3_Crossover}, while in the symmetric limit, $N_\text{G} \sim 1/{N}$. Thus, a symmetric blend, with a large degree of polymerization, has a much smaller critical region than a polymer blend in the asymmetric limit. This fact, combined with the consequences of Eq.~(\ref{Eq_PolyChainFree_PolyBlend}), means that the symmetric polymer blend would only exhibit one type of crossover, namely, the crossover from Ising-like critical behavior in the vicinity of the critical point to mean-field behavior away from the critical point. As the polymer blend becomes asymmetric, a tricritical region would emerge as discussed in Sec.~\ref{Sec3_Crossover}. 


\section{Conclusion \label{Sec_Conclusion}}
We have shown that the nature of critical fluctuations in polymer solutions, including the crossover from Ising-like critical behavior near the critical point of mixing to tricritical behavior near the theta point, has been resolved both theoretically and experimentally. Specifically, this crossover is governed by a competition between two characteristic length scales, namely, the correlation length of concentration fluctuations, $\xi$, related to the osmotic susceptibility, which diverges at the critical point, and the polymer-chain correlation length, $\xi_\psi$, related to the radius of gyration, $R_\text{g}$. This competition causes a coupling between the two order parameters: the conserved order parameter associated with the polymer concentration, $\phi$, and the non-conserved order parameter associated with the fluctuations of the long polymer chains, $\psi(\vec{r})$, which accounts for the observed static and dynamic phenomena in polymer solutions.

\section*{Acknowledgments}
J.V.S. and M.A.A. acknowledge a long-lasting friendship and fruitful interactions with Michael E. Fisher and Benjamin Widom. M.A.A. and T.J.L. also thank Michael Rubinstein, Ralph H. Colby, and Nicholas Tang for recent stimulating discussions on the nature of the crossover from Ising-like criticality to theta-point tricriticality. The research of M.A.A. and T.J.L. was supported by NSF award no. 1856479.

\bibliographystyle{ws-rv-van}
\bibliography{references}

\end{document}